\documentclass[aps,prb,amsmath,amssymb,reprint,showkeys]{revtex4-2}

\usepackage[dvipsnames]{xcolor}
\usepackage{graphicx}
\usepackage{hyperref}
\usepackage[utf8]{inputenc}
\usepackage[T1]{fontenc}
\usepackage{physics}
\usepackage{mathtools}
\usepackage{siunitx}
\usepackage{bm}
\usepackage{dcolumn}
\usepackage{appendix}
\usepackage{lipsum}
\usepackage{multirow}

\usepackage[normalem]{ulem}

\newcommand\T{\rule{0pt}{3.0ex}}
\newcommand\B{\rule[-1.2ex]{0pt}{0pt}}
\newcommand{\ra}[1]{\renewcommand{\arraystretch}{#1}}

\begin{document}
\title{Cluster expansion constructed over Jacobi-Legendre polynomials for accurate force fields}
\author{M.~Domina*}
\author{U.~Patil*}
\author{M.~Cobelli*}
\author{S.~Sanvito}
\affiliation{School of Physics and CRANN Institute, Trinity College, Dublin 2, Ireland}
\date{\today}

\begin{abstract}
We introduce a compact cluster expansion method, constructed over Jacobi 
and Legendre polynomials, to generate highly accurate and flexible 
machine-learning force fields. The constituent many-body contributions are 
separated, interpretable and adaptable to replicate the physical knowledge 
of the system. In fact, the flexibility introduced by the use of the Jacobi
polynomials allows us to impose, in a natural way, constrains and symmetries 
to the cluster expansion. This has the effect of reducing the number of 
parameters needed for the fit and of enforcing desired behaviours of the potential.
For instance, we show that our Jacobi-Legendre cluster expansion can be designed
to generate potentials with a repulsive tail at short inter-atomic distances, 
without the need of imposing any external function. Our method is here continuously
compared with available machine-learning potential schemes, such as the atomic
cluster expansion and potentials built over the bispectrum. As an example we construct a 
Jacobi-Legendre potential for carbon, by training a slim and accurate model capable 
of describing crystalline graphite and diamond, as well as liquid and amorphous 
elemental carbon.
\end{abstract}

\maketitle

\section{Introduction}

Machine-learning potentials (MLPs) are rapidly becoming the gold standard 
for molecular dynamics and thermodynamical sampling in materials
science~\cite{Review1, Review2, Review3, CuDataset}. The general 
idea is that of performing a high-dimension fit of the potential energy 
surface (PES) computed with an electronic {\it ab-initio} method, for instance 
with density functional theory (DFT), to obtain a numerical atomic energy 
functional for large-scale simulations. In particular, one aims at using 
a conveniently limited number of electronic structure data to interpolate 
the potential energy surface at an accuracy comparable with that of the 
electronic method itself. MLPs can thus be defined as parametric functions 
that associate to a given chemical structure the system energy. The mathematical 
relation between the input features describing a structure, often called 
{\it descriptors}, and the output target can be either linear~\cite{SNAP,ACE,MTP} 
or non linear~\cite{NN,GAP,SOAP}. Furthermore, the target quantity may be 
different from the energy and may include electronic 
properties~\cite{Domina,CeriottiNMR,Lunghi2019a,Lunghi2019b}, or even tensorial 
quantities~\cite{HaVuNguyen}.

The specific descriptors choice is crucial to the construction of a MLP. It 
is commonly agreed that a strategy to drastically reduce the size of the training 
set and to improve the model accuracy is that of designing descriptors invariant 
with respect to the symmetries of the target quantity. In the case of the total
energy, this results in descriptors, which are invariant for translations, rotations 
and permutations of identical atoms. In principle, one can then combine any choice of 
descriptors with any desired machine-learning model, going from simple regressions, 
to neural networks of various complexity, to kernel-based schemes. Typically, 
there is a subtle tradeoff between the model complexity, the descriptor type and 
the size and composition of the dataset needed to construct the MLP. 
Complex many-body descriptors~\cite{Musil2021} are usually combined with linear 
models, while simpler structure representations are used as input to deep-learning
algorithms. In both cases, there may be issues of interpretability, namely it is 
not always transparent what is the level of physics learned by the model itself. 
As a consequence one often relies on numerical techniques to establish whether a 
particular atomic configuration is interpolated or extrapolated by the 
model~\cite{MaxVol}.

In this work we introduce a novel linear model built over a set of descriptors
derived from the energy cluster expansion. Our MLP, that we name the Jacobi-Legendre 
Potential (JLP), is close in spirit to the recently introduced Atomic Cluster 
Expansion (ACE)~\cite{ACE,PACE}. In fact, given the completeness of the 
ACE~\cite{ACECompl}, one can establish a one-to-one mapping between the 
two potentials. Importantly, our JLPs adopt internal coordinates, so that they 
are, in essence, expansions of the $N$-body potentials in orthogonal polynomials 
evaluated on distances and angles between atoms. As such, the JLPs are not 
affected by issues concerning the invariant coupling of different angular momenta 
channels \cite{GAP1,Bartok}. Our use of the internal coordinates is closer 
to the recently developed Proper Orthogonal Descriptors (PODs) \cite{POD1,POD2}. 
Here, however, we retain the spherical harmonics formalism by mean of the Legendre
polynomials, so that a comparison between the JLPs and the other well-known 
potentials can be naturally drawn. Our scheme also makes extensive use of Jacobi
polynomials, of which the Legendre ones are a particular case.

One of the most important property of MLPs is the achievement of linear scaling
with respect to the number of atoms in the neighborhood of a chosen one. Here, we 
will show that linear scaling can be achieved for the JLPs too and, in doing so, 
we will establish a link between well-known MLPs and the internal coordinate 
representation used here.

The use of an explicit expansion over orthogonal and complete polynomials gives 
several advantages, such as the enforcement of symmetries and local constraints. 
Indeed, our potentials are constructed so that key properties, such the smooth
vanishing contribution at the cut-off radius, arise naturally without the need of
introducing ad-hoc cut-off functions. In fact, these properties are enforced by
applying constrains on the expansion coefficients. Crucially, the procedure here
presented is completely general, so that not only the number of coefficients to 
learn can be substantially reduced, but also the physical knowledge of the 
PESs can be introduced, in a natural way. 

For instance, a desired feature arising from the choice of the Jacobi polynomials
and of the constraining procedure, is that, by appropriately tuning the 
hyper-parameters, a repulsive behaviour naturally emerges for the potential at small 
distances. This is also obtained without introducing any external repulsive function. 
Moreover, while it is not generally possible to completely separate the body-order 
contributions, we formally avoid any mixing between them. This allows us to 
reconstruct the $N$-body functional dependence in terms of the learned coefficients. 
As a consequence, by combining these two properties, one can introduce an 
inductive-bias in the models by selecting, for example, only the hyper-parameters 
that lead to a repulsive short-range behaviour of the two-body interaction. 
Since small distances are usually absent from the training set, a direct consequence
is that the potentials naturally possess a physically meaningful behaviour in this 
extrapolated regime. 

The paper is organized as follows. An extensive Methods section presents in 
detail each body-order of the expansion, with a discussion on the relevant 
properties of each term. Then, the potential is fitted to the carbon dataset used 
to train the GAP17 potential of Reference~\cite{gapC}. The result of the 
fit on energies, forces and stress are reported. Furthermore, we will close 
this case study by presenting the phonon dispersion curves for graphene 
and diamond predicted by the trained JLP model.

\begin{figure*}
\includegraphics[width=0.9\textwidth]{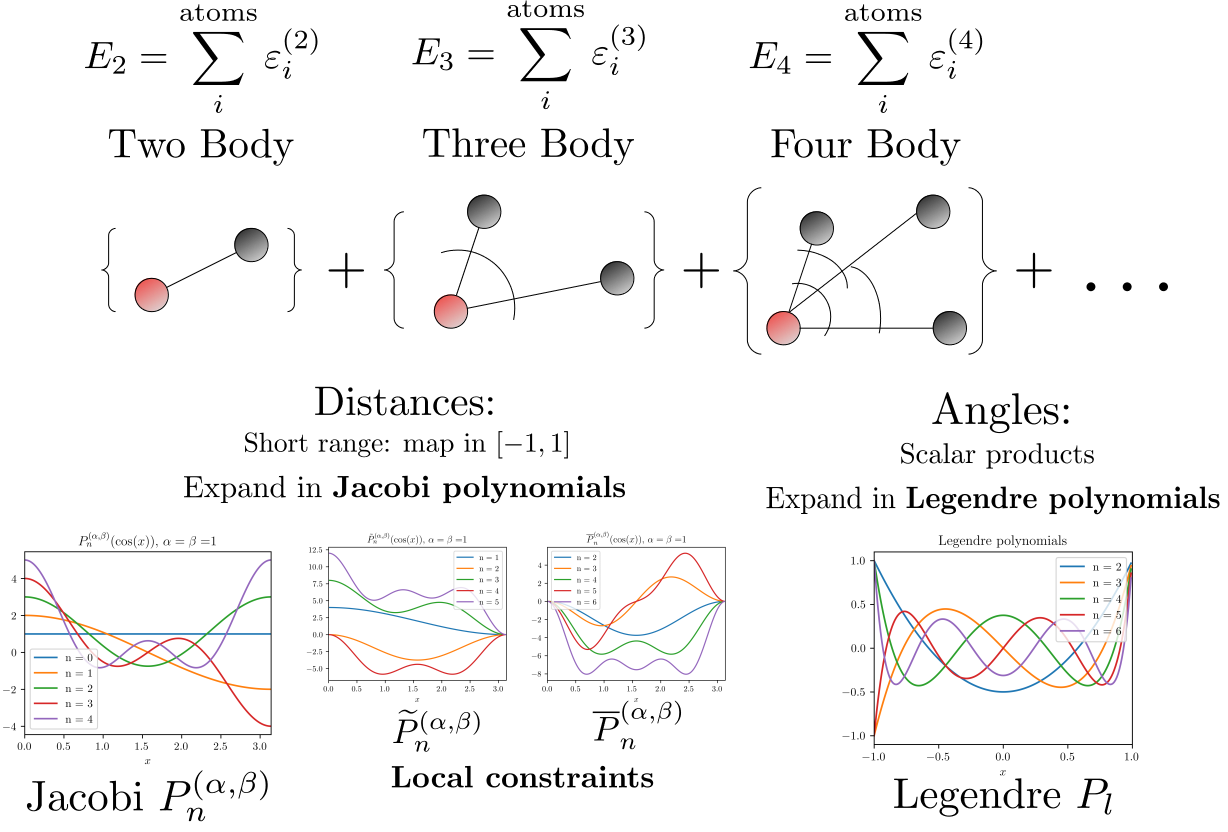}
\caption{The workflow of the linear model presented in this work. We first 
decompose the total energy over the terms of a local multi-body expansion, as 
in Eq.~\eqref{eq:1}. Each contribution is then further expanded over atomic
contributions [Eq.~\eqref{eq:2}], $\varepsilon_i^{(n)}$, which depend on
the distances and the angles between a central atom (in red) and atoms in
its neighborhood (in black). For instance, the two-body term consists only of 
one distance, the three-body one of two distances and one angle, etc. By assuming 
short-range interaction, the distances are then mapped onto the interval 
$[-1,1]$, so that each distance-dependent term can be expanded as products 
of Jacobi polynomials. The angles are then mapped onto scalar products, so 
that the functional dependence on the angles can be similarly expanded in 
terms of Legendre polynomials. Crucially, the expansions on the distances 
is locally constrained, so that effective polynomials will be employed in 
place of the Jacobi polynomials.}
\label{fig:1}
\end{figure*}

\section{Methods}
In this section, we introduce the JLPs. This class of potentials is based on 
the total energy cluster (many-body) expansion. Therefore, after a discussion 
of the main idea behind such strategy, we will proceed with the systematic
introduction of each many-body term and their associated technical details. 
Note that a similar strategy can be also used to construct JLP-like model 
for quantities different from the energy, such as the charge density at a
particular point in space~\cite{Bruno}

\subsection{Introduction}
An overview of the strategy behind the construction of the JLPs is provided in
Fig.~\ref{fig:1}. In general, it is reasonable to assume that a system total 
energy, $E$, can be partitioned into a short- and a long-range contribution. 
Our proposed MLP accounts only for short-ranged part, $E_\text{short}$, that 
can be further expanded over terms vanishing at distances larger than a 
characteristic interaction range. In particular we follow the well-known strategy 
of a multi-body expansion for the energy and write
\begin{equation}\label{eq:1}
E_\text{short} = E_1 + E_2 + E_3 + E_4\ldots\:.
\end{equation}
Here the single-body contribution, $E_1$, is an energy offset depending on 
the number of atomic species present in the system, $E_2$ is the two-body (2B)
energy, depending only on atoms pairs, $E_3$ is the three-body (3B) energy depending
on triplets of atoms and, in general, $E_n$ describes the $n$-body ($n$B) energy 
term. 

A second essential assumption is that the we can decompose each of the $n$B 
energy term in local quantities, such that each term can be written as a sum 
of atom-centered contributions. Explicitly, this writes
\begin{equation}\label{eq:2}
    E_n = \sum_i^\text{atoms} \varepsilon_i^{(n)}\:,
\end{equation}
with $n\ge 2$, and where the sum runs over all possible atoms in the system. 
Each local contribution to the $n$B energy, $\varepsilon_i^{(n)}$, depends only 
on the local neighborhood of the $i$-th atom (the red atom in Fig.~\ref{fig:1}), 
up to a cut-off distance $r_\text{cut}$.

In essence, the JLPs consist of a linear expansion of the $\varepsilon_i^{(n)}$
contributions. As such, at the core, the JLP is closely related to linear MLPs 
such as the Spectral Neighbour Analysis Potential (SNAP)\cite{SNAP}, the Moment 
Tensor Potentials (MTPs) \cite{MTP}, and the Atomic Cluster Expansion (ACE) 
\cite{ACE,PACE}. Since the successful generalization of the coupling scheme of 
the powerspectrum (a 3B representation) and the bispectrum (a 4B representation)
\cite{GAP1,Bartok} to any higher-body order, first introduced in the ACE potentials, 
all new potentials build from the same set of assumptions (many-body expansion of 
the energy and locality), differ in the way of constructing the basis functions, 
or on the introduction of completely new basis sets \cite{Review1,Review2}. The 
JLPs are not different in this regards. Based on a particular choice of basis
functions (radial and angular), they are also complete, so that a one-to-one 
mapping between the terms of a JLP and the analogous ones of the ACE is possible. 
In particular, as the name suggests, we chose the Jacobi polynomials as radial 
basis and the Legendre polynomials as the angular one. 

The choice of Jacobi polynomials \cite{Abramovitz}, $P^{(\alpha,\beta)}_n(x)$, 
is motivated by their dependence on the two real parameters, 
$\alpha$ and $\beta$, which can 
lead to a broad selection of different orthogonal polynomials. Two classical
examples are the Legendre polynomials ($\alpha= \beta = 0$) and the Chebyshev
polynomials of the second kind ($\alpha= \beta = 1/2$). Thus, treating
$\alpha$ and $\beta$ as hyper-parameters allows one to optimize the radial 
basis set, and removes the need for manually choosing the best basis. 
In contrast, we have chosen the Legendre polynomials not only since 
they lead to a certain homogeneity in the representation (being the Legendre
polynomials a particular instance of the Jacobi ones), but also for their strong
relation with the spherical harmonics. This means that a spherical harmonics
decomposition can always be performed, a key feature for achieving 
computational-linear scaling with respect to the number of atoms (neighbours) 
inside the interaction cut-off sphere. 

After performing the expansion over the chosen basis, we will present a general 
way for constraining the expansion coefficients, so that known physical (and local) 
properties of the system can be encoded directly in the descriptors at any body 
order. As a bi-product of applying the constrain on Jacobi polynomials, we will 
show the natural emergence of the widely-used cut-off function, $f_c = 
(1-\cos(x))/2$. As far as we know, this is the only case in which a cut-off
function, $f_c$, is not externally imposed on the basis set, but instead 
emerges naturally from the formalism. 

Finally, since it has been proved that all four-body descriptors mentioned before 
are not complete, in the sense that one could find two distinct local environments
with the same set of descriptors \cite{Incompleteness}, or manifold with slow-varying 
fingerprints with respect to a similarity measure 
\cite{DistMeasure, QuasiConsManifold}, we will explicitly 
investigate the JLP up to the five-body order term, $E_5$. We will then 
briefly discuss that the internal coordinates constitute an over-complete set at 
the five-body level, so that the coupling introduced in ACE is preferred to the 
one presented here. However, we anticipate that the choice of the Jacobi polynomial
as a basis set, and the associated constraining procedure, can be applied also to
other potentials. Indeed, they can be exported easily to other multi-body expansion
approaches, so that, for example, one could use the constrained-Jacobi basis as a 
radial basis for ACE.

\begin{figure}
\includegraphics[width=0.9\columnwidth]{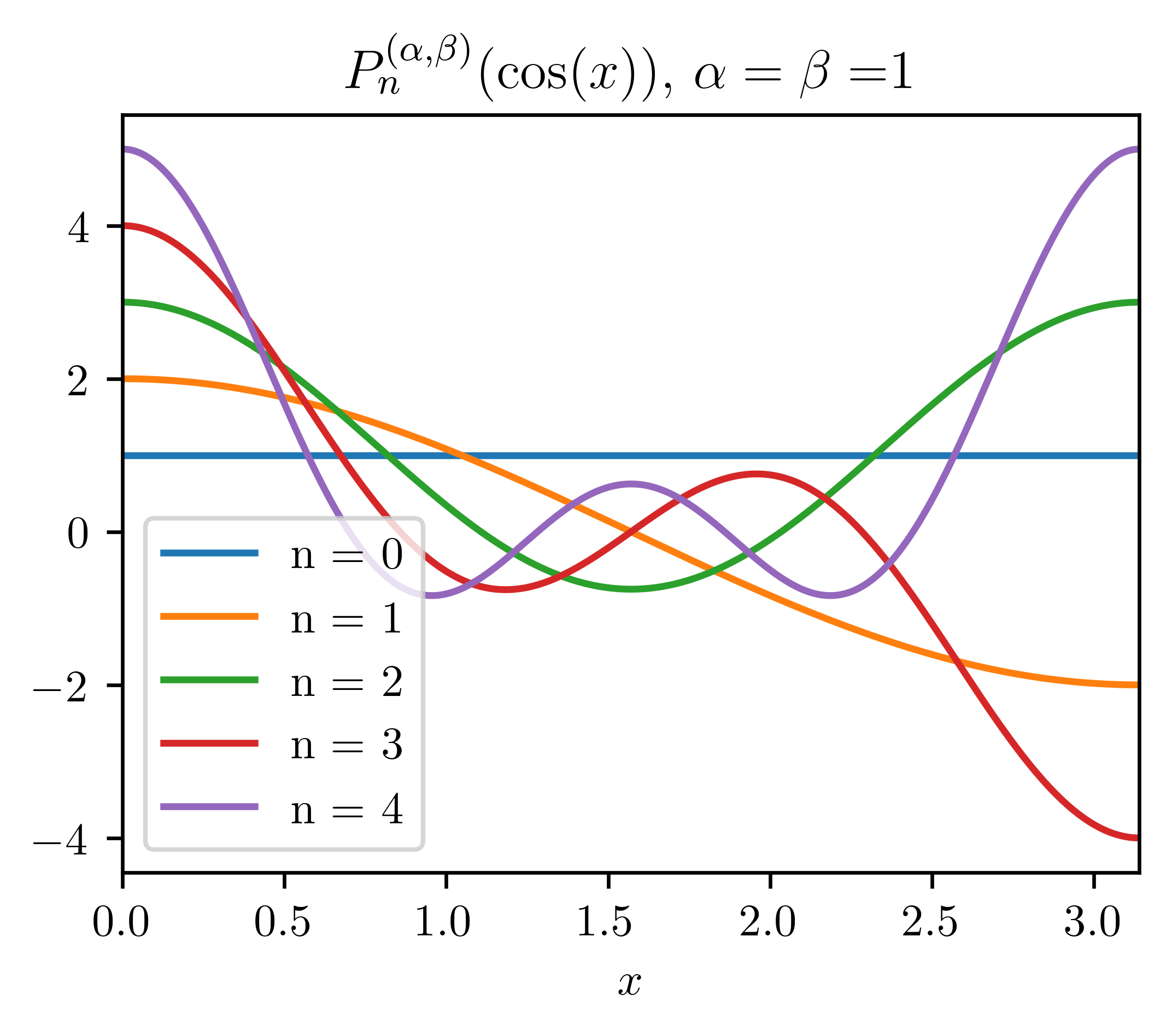}
\caption{First five Jacobi polynomials, $P_n^{(\alpha,\beta)}$, with 
$\alpha = \beta = 1$. In the plot the functions are composed with a cosine, 
so to give a better idea of the representation used in the expansion of 
Eq.~\eqref{eq:8}. It can be appreciated how the derivative is zero at both edges 
of the domain.}
\label{fig:2}
\end{figure}
\subsection{The Two-Body Term}
In this section we introduce the expansion of the two-body energy term, $E_2$. 
Since the total energy is a scalar, it must be invariant under translations 
and rotations of the reference frame. A possible way to satisfy such invariance
is to assume that the energy depends only on the distances between atom pairs, 
and that this dependency is realized by a continuous function (potential),
$v^{(2)}$. Furthermore, we assume that the actual functional form depends only 
on the atomic species of the atoms involved, so that, if $Z_i$ is the atomic number
of the atom located at the position $\mathbf{r}_i$, and 
$r_{ji}=|\mathbf{r}_j-\mathbf{r}_i|$, we have 
\begin{equation}
    v^{(2)}\equiv v^{(2)}(r_{ji};Z_j,Z_i) \equiv v_{Z_j Z_i}^{(2)}(r_{ji}),
\end{equation}
and
\begin{equation}\label{eq:4}
E_2 = \sum_{\substack{ij\\j\neq i}} v^{(2)}_{Z_j Z_i}(r_{ji})\:.
\end{equation}
The two-body potentials, $v_{Z_j Z_i}^{(2)}$, is thus defined symmetric 
under the exchange $Z_j \leftrightarrow Z_i$, namely 
$v_{Z_j Z_i}^{(2)} = v_{Z_i Z_j}^{(2)}$. 
Note that, in principle, one can still explicitly 
distinguish nonequivalent atoms belonging to the same specie, by 
introducing ``virtual'' species.

It is useful to remark that the 2B term in Eq.~\eqref{eq:4} can be re-casted 
in the form of Eq.~\eqref{eq:2}, where 
$\varepsilon^{(2)}_i=\sum_{j\neq i}v^{(2)}_{Z_j Z_i}(r_{ji})$ is
the energy associated to the $i$-th atom resulting from the pairwise 
interaction with its local atomic neighbourhood. Note that that these local 
contributions are well defined because of the short-ranged nature of the 
interaction. Thus, there exists a natural cut-off radius $r_\text{cut}$, 
such that $v_{Z_j Z_i}(r_{ji})\simeq 0$ for $r_{ji}>r_\text{cut}$. 

\begin{figure}
\includegraphics[width=0.9\columnwidth]{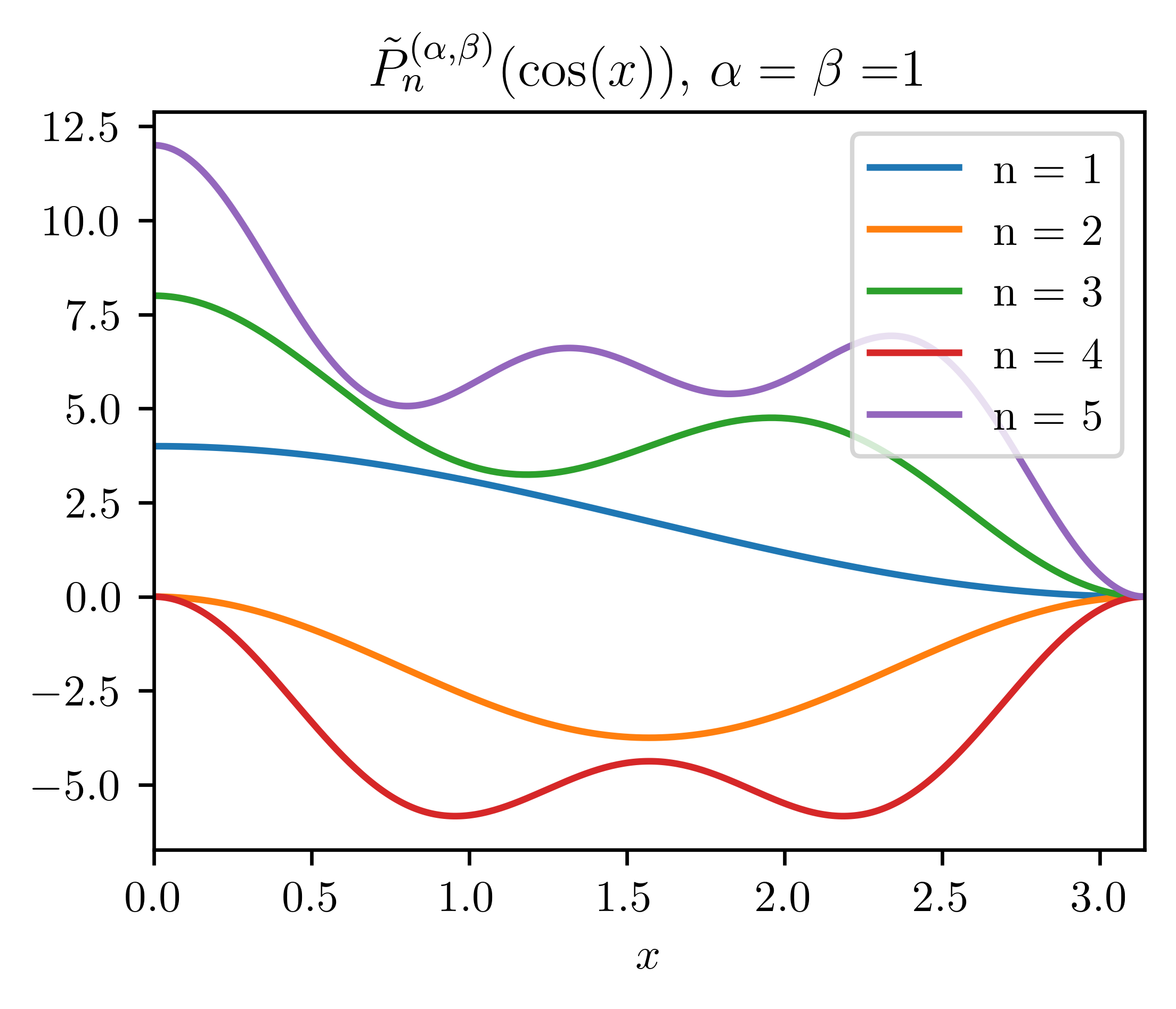}
\caption{First five vanishing-Jacobi polynomials, 
$\widetilde{P}_n^{(\alpha,\beta)}$ ($\alpha=\beta=1$) as defined 
in Eq.~\eqref{eq:6}, derived from the Jacobi polynomials of Fig. \ref{fig:2}. 
These polynomials are constrained to vanish at the right-hand side edge of 
the domain.}
\label{fig:3}
\end{figure}

We now provide the proposed expansion for the potentials $v_{Z_j Z_i}^{(2)}$, followed by its derivation. The expansion is
\begin{equation}\label{eq:5}
v^{(2)}_{Z_j Z_i}(r_{ji}) = \sum_{n=1}^{n_\text{max}} a_n^{Z_j Z_i} \widetilde{P}_n^{(\alpha,\beta)}\left(\cos\left(\pi\dfrac{r_{ji}-r_\text{min}}{r_\text{cut}-r_\text{min}}\right)\right)\:,
\end{equation}
where the sum is truncated to a suitable polynomial order, $n_\text{max}$,
and where $a_n^{Z_j Z_i}$ are the expansion coefficients for the $n$-th order. 
The vanishing-Jacobi polynomials, $\widetilde{P}_n^{(\alpha,\beta)}$, 
employed here are defined in terms of the Jacobi polynomials, 
$P_n^{(\alpha,\beta)}$, as
\begin{equation}\label{eq:6}
\widetilde{P}_n^{(\alpha,\beta)}(x) = {P}_n^{(\alpha,\beta)}(x) - P_n^{(\alpha,\beta)}(-1)\quad\text{for }-1\leq x \leq 1\:,
\end{equation}
for $n\ge 1$. Thus, the $\widetilde{P}_n^{(\alpha,\beta)}$ have the 
property to vanish at the right-hand side extreme of their domain, 
namely at $r_\mathrm{cut}$.
The Jacobi polynomials are shown in Fig.~\eqref{fig:2}, while 
the corresponding vanishing-Jacobi polynomials are in Fig.~\ref{fig:3}. 
The expansion presents five hyper-parameters, $\alpha$, $\beta$, 
$r_\text{cut}$, $r_\text{min}$ and $n_\text{max}$, with $\alpha$ and $\beta$ 
being real numbers greater than $-1$. We will refer to Eq.~\eqref{eq:5} as the 
2B-Jacobi-Legendre (2B-JL) expansion.

At this point, it should be noted that there can be a different set of 
hyper-parameters for each different atomic species. As such, in the case of 
many-species compounds the hyper-parameter space can potentially become 
rather large. Then, it may be desirable to take system-based approximations 
or to perform features selection.

We will now present the arguments leading to Eq.~\eqref{eq:5}. In order 
to make the formalism more readable, we define the compact notation
\begin{equation}\label{eq:7}
    \widetilde{P}_{nji}^{(\alpha,\beta)} \equiv \widetilde{P}_n^{(\alpha,\beta)}\left(\cos\left(\pi\dfrac{r_{ji}-r_\text{min}}{r_\text{cut}-r_\text{min}}\right)\right) \:,
\end{equation}
which will be widely used throughout this work. 
As already remarked, the potentials should vanish for distances larger 
than the interaction cut-off radius. With only this constrain in mind, 
we can expand the potential in terms of Jacobi polynomials as
\begin{equation}\label{eq:8}
    v^{(2)}(r) = \sum_{n=0}^{n_\text{max}} a_n
    P_n^{(\alpha,\beta)}\left(\cos\left(\pi r/r_\text{cut}\right)\right)\: ,
\end{equation}
where, for simplicity, we set $r_\text{min} = 0$ and omit the explicit 
dependence on the atomic species.

We have chosen the Jacobi polynomials, $P_n^{(\alpha,\beta)}(x)$, since 
they are complete and orthogonal over the interval $x \in [-1,1]$. 
Furthermore, as already noted, their generality, parameterised through the
real coefficients $\alpha$ and $\beta$, allows one to perform  
automatic searches of the most efficient basis set, without any additional 
hypothesis. We found that, in most cases, there is a large range of 
optimal $\alpha$ and $\beta$ values, so that we usually reduce the number 
of hyper-parameters by constraining the search to $\alpha = \beta $.

Note that we are not introducing any cut-off function in the expanson 
to force a smooth-vanishing behaviour at the cut-off radius. Also, the 
$a_0$ coefficient is not present in the sum of Eq.~\eqref{eq:5}, while 
still appears in Eq.~\eqref{eq:8}. We will now impose the right behaviour 
on the expansion coefficients, so that the resulting potential vanishes 
by construction at the cut-off radius. The result of this approach is
similar, in a sense, to models with naturally vanishing radial functions, 
such as the Spherical-Bessel descriptors~\cite{SphericalBessel}. 
Explicitly, we constrain the expression in Eq.~\eqref{eq:8} to satisfy 
the condition $v^{(2)}(r_\text{cut}) = 0$. Then, 
since $P^{(\alpha,\beta)}_0 = 1$, we obtain that the first coefficient 
must satisfy
\begin{equation}\label{eq:9}
    a_0 = - \sum_{n\geq 1}^{n_\text{max}}a_n P_n^{(\alpha,\beta)}(-1)\: .
\end{equation}
By inserting this expression back into Eq.~\eqref{eq:8}, we finally obtain 
Eq.~\eqref{eq:5}. It is worth mentioning that this procedure can be easily
generalized to impose any constrain to the functional form of the potential,
so that local physical knowledge of the system can be enforced in the 
description itself. An example of a further constrain will be shown 
for higher-body terms. 

We can then interpret the vanishing Jacobi polynomials, defined in 
Eq.~\eqref{eq:6}, as the radial basis obtained when expanding functions 
vanishing at the left-hand side limit of the interval $[-1,1]$ (the point 
$x=-1$ is mapped onto the cut-off distance in our representation). As a 
final remark, the expansion coefficients $a_n^{Z_j Z_i}$ inherit the same 
symmetry properties of the potential, namely they are symmetric under the 
exchange of the atomic species, $a_n^{Z_j Z_i} = a_n^{Z_i Z_j}$. 

In closing this section, it must be mentioned that the 2B-JL expansion
suffers from the same scaling problem of most of the established MLPs 
when dealing with multiple species. In fact, the number of pair-wise 
potentials that one can define scales quadratically with the number of
species, so that system-based approximations are required for complex chemical
compositions. This problem will become more severe for the higher-body 
terms. 

\subsubsection*{Emergence of the cut-off function from the constrains}

A relevant property of the 2B-JL expansion is that, as rigorously proved in
Appendix \ref{app:A}, we can factorize the vanishing Jacobi polynomials as
\begin{equation}\label{eq:10}
    \widetilde{P}_n^{(\alpha,\beta)}(\cos x) = f_c(x) Q^{(\alpha,\beta)}_n(\cos(x))\: ,
\end{equation}
where $f_c(x)$ is the well-know cut-off function $f_c(x) = (1+\cos(x))/2$, 
first introduced in reference \cite{symmfun}, and the 
$Q_n^{(\alpha,\beta)}(\cos(x))$ are functions explicitly defined in 
Appendix \ref{app:A}. As far as we know, the functions $Q_n^{(\alpha,\beta)}$ 
are not equivalent to other functions already used in the MLPs literature. 
While the property described by Eq.~\eqref{eq:10} establishes a strong
connection between our expansion and other potentials, which use the cut-off 
function, it is important to stress that with the JLPs $f_c$ arises naturally from 
the choice of the radial basis and the constraining method implemented. As such, 
it is not an embedding function, as one can clearly see in Fig. \ref{fig:3}. 
Among the advantages of this approach there is that, since the Jacobi 
polynomials are already complete and orthogonal, no further orthogonalization
procedure has to take place. Also, we do not have to explicitly evaluate the 
derivative of the cut-off function when calculating the forces, since we can
simply use the derivative of the (vanishing-) Jacobi polynomials. Finally, by
imposing the constrain of Eq.~\eqref{eq:9}, we are reducing the number of 
coefficients to learn: this is particularly relevant for higher-body terms, 
as it will be shown in the following sections.

\subsection{The Three-Body Term}
In this section we will discuss the linear expansion of the three-body 
energy term, $E_3$. While the core strategy is the same as the one employed 
in the previous section, for $E_3$ here we will introduce a Legendre expansion
for the angular dependence of the cluster, we will impose a further constrain
on the Jacobi polynomials and we will discuss the role of symmetries when 
considering atoms of the same species. 

Following the same approach introduced in the previous section, we assume 
that $E_3$ can be written as a sum of local 3B potentials, $v^{(3)}$, as 
\begin{equation}\label{eq:11}
E_3 = \sum_{i}^\text{atoms} \sum_{(j,k)_i}v^{(3)}_{Z_j Z_k Z_i}(r_{ji},r_{ki},\hat{\bm r}_{ji}\cdot \hat{\bm r}_{ki})\:,
\end{equation}
where the first sum runs over all the atoms in the system and the second one
runs over all the atoms pairs in the neighbourhood (within $r_\mathrm{cut}$) 
of the $i$-th atom (the red atom in Fig.~\ref{fig:1}). Here, in order to ensure 
the translational and rotational invariance of the descriptors, we consider
only internal coordinates between the central atom $i$ and the atoms $j$ and
$k$ in the surroundings. Therefore, only the distances $r_{ji}$ and $r_{ki}$
and the scalar products $\hat{\bm r}_{ji}\cdot \hat{\bm r}_{ki}$ (essentially 
the angle defining a three-body cluster), are taken into consideration.

The functional form of the potential $v^{(3)}_{Z_j Z_k Z_i}$ depends on the
ordering of the atomic species numbers $Z_j$, $Z_k$ and $Z_i$. Specifically,
the first atomic species refers to the first distances, the second atomic
species to the second distance, while the last one refers to the central 
atom. Thus, it holds that
\begin{equation}\label{eq:12}
    v^{(3)}_{Z_k Z_j Z_i}(r_{ki},r_{ji},s_{jki}) = v^{(3)}_{Z_j Z_k Z_i}(r_{ji},r_{ki},s_{jki})\:,
\end{equation}
where $s_{jki}$ is a short-hand notation for 
$\hat{\bm r}_{ji}\cdot \hat{\bm r}_{ki}$. Put it in words, if we exchange the 
species of the atoms in the environment, we will also have to exchange their
distances. From now on, we will use $v_{jki}^{(3)}$ as a shorthand notation
for $v^{(3)}_{Z_j Z_k Z_i}$.
\begin{figure}
\includegraphics[width=0.9\columnwidth]{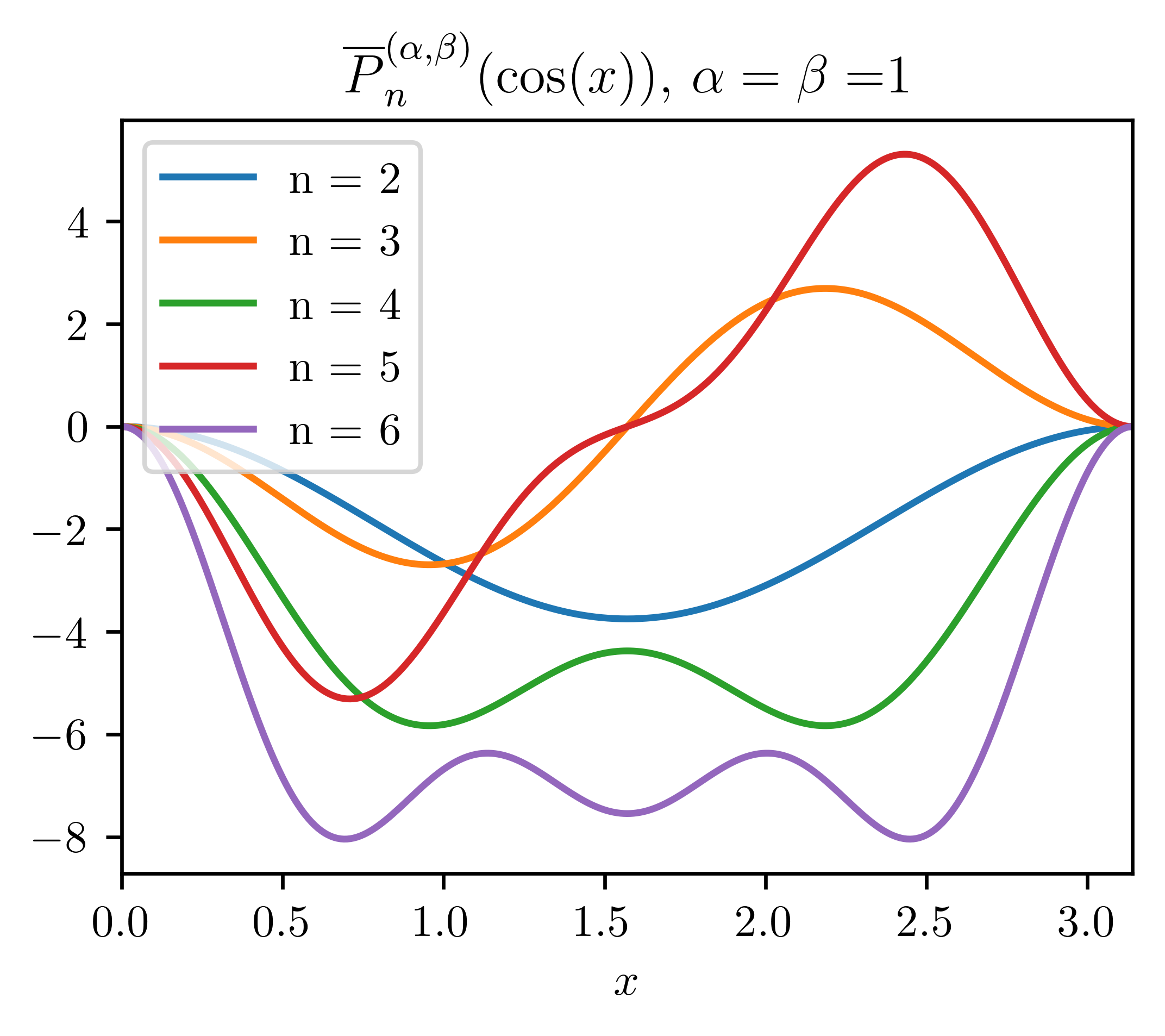}
\caption{The first five double-vanishing Jacobi polynomials,
$\overline{P}_{n}^{(\alpha,\beta)}$ (here plotted for $\alpha=\beta=1$) 
as defined in Eq.~\eqref{eq:13} derived from the vanishing ones shown 
in Fig. \ref{fig:3}. The polynomials are constrained to vanish at both 
edges of the domain.}
\label{fig:4}
\end{figure}

By adopting the same workflow followed in constructing the 2B case, we 
now give an expression for the 3B JL expansion, and then we provide its 
derivation. The 3B-JL expansion reads
\begin{equation}\label{eq:13}
v^{(3)}_{jki}(r_{ji},r_{ki},s_{jki}) = \sum_{n_1, n_2 = 2}^{n_\text{max}}\sum_{l=0}^{l_\text{max}} a_{n_1 n_2 l}^{jki} {\overline{P}}_{n_1 ji}^{(\alpha,\beta)}\overline{P}_{n_2ki}^{(\alpha,\beta)}P_l^{jki}\:,
\end{equation}
where $P_l^{jki} = P_l(s_{jki})$ is the Legendre Polynomial, $P_l$, evaluated 
on the scalar product $s_{jki}$. The sum runs on all the $n_1$ and $n_2$ in the
interval [2,$n_\text{max}$]. The 3B expansion introduces a new hyperparameter, $l_\text{max}$, which sets the level of truncation of the angular expansion. 
The coefficients $a^{j k i}_{n_1 n_2 l}$ have to be intended as a compact 
form for $a^{Z_j Z_k Z_i}_{n_1 n_2 l}$. Crucially, we use here the 
double-vanishing Jacobi polynomials, $\overline{P}_{n}^{(\alpha,\beta)}(x)$, 
which can be defined in terms of the vanishing ones as (see Fig.~\ref{fig:4}) 
\begin{equation}
\overline{P}_{n}^{(\alpha,\beta)}(x) = \widetilde{P}_{n}^{(\alpha,\beta)}(x) - \dfrac{\widetilde{P}_{n}^{(\alpha,\beta)}(1)}{\widetilde{P}_{1}^{(\alpha,\beta)}(1)}\widetilde{P}_{1}^{(\alpha,\beta)}(x) \;,
\end{equation}
for $n\ge 2$. From this definition it can be seen that the double-vanishing
polynomials not only vanish smoothly at the cut-off distance ($x = -1$) but 
also for small distances ($x = 1$). By employing these polynomials, the repulsive
behaviour at short distances is not influenced by the 3B-JL expansion and, as 
such, is completely determined by the 2B expansion. Note that these polynomials 
has been devised for the case in which $r_\text{min}$ is small. If this hypothesis
does not hold, we suggest a case-by-case investigation of the most appropriate 
polynomials or constraints to use.

The derivation of the expansion in Eq.~\eqref{eq:13} follows the same 
strategy presented in detail for the derivation of the 2B-JL expansion, 
Eq.~\eqref{eq:5}. Since the distances and the scalar product are independent
variables, we expand the functional dependence of the potential on the distances 
in terms of a product of two Jacobi polynomials, one for each distance. Then,
the scalar product dependence is expanded in terms of Legendre polynomials.
Analogously to the constrain adopted in the 2B case, we constrain the expansion 
to vanish at the cut-off radius. Here, however, we impose the potential to
vanish when \emph{at least one} of the distances approaches the cut-off, 
independently of the value of the other distance or of the angular part. 
Crucially, applying independent constrains on the variables at play, allows 
us to severely reduce the number of free coefficients, when compared to the 
2B case. Indeed, the constrains explicitly read [please, compare with the
constrain introduced in Eq.~\eqref{eq:9}]
\begin{equation}
    \nonumber
    \begin{dcases}
    a_{0 n_2 l} = -\sum_{n_1\geq 1}^{n_\text{max}} a_{n_1 n_2 l}P_{n_1}^{(\alpha,\beta)}(-1)\qquad\text{for all  }n_2,l,\\
    a_{n_1 0 l} = -\sum_{n_2\geq 1}^{n_\text{max}} a_{n_1 n_2 l}P_{n_2}^{(\alpha,\beta)}(-1)\qquad\text{for all  }n_1,l,\\
    \end{dcases}
\end{equation}
so that the expression can be re-casted in terms of products of vanishing 
Jacobi polynomials and Legendre polynomials only. However, we can further
constrain the number of free coefficients by imposing that the potentials 
also vanish when at least one of the distances approaches zero. In this way 
we impose a condition also on the $a^{j k i}_{1 n_2 l}$ and $a^{j k i}_{n_1 1 l}$
coefficients. In doing so, we obtain the double-vanishing polynomials and 
the 3B-JL expansion of Eq.~\eqref{eq:13}. Note that in the unconstrained 
case we have $(n_\text{max}+1)^2(l_\text{max}+1)$ free coefficients, while 
in the double-constrained one these are only $(n_\text{max}-1)^2(l_\text{max}+1)$. As such, we deduce that the reduction in the number of coefficients is quite
severe for relatively low $n_\text{max}$. Another relevant reduction in the 
number of free parameters is induced through the symmetries of the coefficients, 
when atoms of the same species are taken into account, as explained in detail 
in the following section. 

\subsubsection*{Symmetries of the coefficients}

We can explicitly read the role of the indexes in the expansion coefficients
$a_{n_1n_2l}^{jki}$ of Eq.~\eqref{eq:13}, by noticing that the first index, 
$n_1$, refers to the expansion on the first argument of the potential 
$v^{(3)}_{jki}$ (the distance between the atoms $j$ and $i$), while the 
second one expands the second distance. Thus, the symmetry property 
of the potentials described by Eq.~\eqref{eq:12} directly implies that 
$a^{j k i}_{n_1 n_2 l} = a^{k j i}_{n_2 n_1 l}$, namely that the 
expansion coefficients are symmetric for simultaneous exchange of the 
species indexes $Z_j\leftrightarrow Z_k$ and of the Jacobi indexes, 
$n_1\leftrightarrow n_2$. While this is effectively just a reordering of 
the arguments of the potential, with appropriate re-labelling, it becomes 
relevant in the case of identical atoms. Indeed, if the atoms $j$ and $k$ 
belong to the same atomic species $Z$, then they are indistinguishable, 
making the potential invariant under the exchange of the first and the second
argument (the two distances). Then, one needs to enforce the same symmetry 
on the coefficients, namely they must be symmetric under Jacobi-index 
exchange alone, $a^{Z Z Z_i}_{n_1 n_2 l}$ = $a^{Z Z Z_i}_{n_2 n_1 l}$. 

We can then re-cast the 3B-JL expansion for the same atom species, 
$Z_j = Z_k = Z$, as
\begin{eqnarray}\label{eq:15}
    &&v^{(3)}_{Z Z Z_i}(r_{ji},r_{ki},s_{jki}) =\nonumber \\
     &&= \sum_{n_1=2}^{n_\text{max}} \sum_{l=0}^{l_\text{max}} a_{n_1 n_1 l}^{Z Z Z_i} {\overline{P}}_{n_1 ji}^{(\alpha,\beta)}\overline{P}_{n_1ki}^{(\alpha,\beta)}P_l^{jki}+\\&&+\sum_{\substack{n_1 =2\\n_2 = 2\\n_1> n_2}}^{n_\text{max}} \sum_{l=0}^{l_\text{max}} a_{n_1 n_2 l}^{Z_j Z_k Z_i} \bigg[{\overline{P}}_{n_1 ji}^{(\alpha,\beta)}\overline{P}_{n_2ki}^{(\alpha,\beta)}+{\overline{P}}_{n_2 ji}^{(\alpha,\beta)}\overline{P}_{n_1ki}^{(\alpha,\beta)}\bigg]P_l^{jki}\:.\nonumber
\end{eqnarray}
Equation~\eqref{eq:15} explicitly shows the application of the symmetries 
for $n_1 \neq n_2$. Now, we can introduce the more practical expression
\begin{eqnarray}\label{eq:16}
    &&v^{(3)}_{Z_j Z_k Z_i}(r_{ji},r_{ki},s_{jki}) =\nonumber \\
    &&= \sum_{n_1 n_2 l}^{\text{unique}} a_{n_1 n_2 l}^{Z_j Z_k Z_i} \sum_{\text{symm.}}\bigg({\overline{P}}_{n_1 ji}^{(\alpha,\beta)}\overline{P}_{n_2ki}^{(\alpha,\beta)}P_l^{jki}\bigg)\:,
\end{eqnarray}
which encompasses also the cases for different species and it is easily
generalized to higher-body order expansion terms. Here, the first sum runs 
over indexes that lead to non-equivalent coefficients with respect to the 
symmetries of the potential (in this case, indexes such that $n_1\ge n_2$), 
while the second sum runs over all the permutations of indexes that refers 
to equivalent coefficients (in this case the exchange $n_1 \leftrightarrow n_2$).
If the atoms $j$ and $k$ belong to two different species, then the expression
reduces to the simple form of Eq.~\eqref{eq:13}. In contrast, if the $j$-th 
and $k$-th atoms are of the same species, we end up with the formula in
Eq.~\eqref{eq:15}. It must be noted that, not only this expression is 
crucial to enforce the r\^ole of identical atoms, but it also roughly halve 
the number of free coefficients in the expansion. Finally, we conclude by noting 
that, while in the case of the 3B expansion, there is no difference between 
the symmetrization in Eq.~\eqref{eq:16} and the lexicographic order introduced 
for the ACE coefficients (see Reference \cite{CompletenessACE} for details), 
these are indeed different in the generalization to the 4B case, as will be 
shown in Section~\ref{sec:fourb}. 

\subsection{Linear Scaling and the JL Atomic Basis}

Before presenting the 4B-JL expansion, we discuss here the scaling of the 
3B-JL expansion, with respect to the number of neighbours inside the cut-off
volume. Indeed, by inserting Eq.~\eqref{eq:16} into the expression of
Eq.~\eqref{eq:11} for the 3B energy, $E_3$, it is clear that the computational
time to evaluate the 3B-JL expansion scales quadratically with the number of
neighbours surrounding a central atom. This is because one has to explicitly 
look around for all the possible pairs of atoms. Such scaling makes the 
formalism unpractical, when the number of atoms inside the cut-off sphere 
becomes relatively large. Most of the MLPs used in literature have solved 
this problem by achieving linear scaling with respect to the number of 
neighbors. Importantly, also the 3B-JL expansion, being strictly tied to 
the powerspectrum components \cite{Bartok}, can be rearranged so to reach 
the same scaling. In this rather technical section, we will mainly discuss 
the results of such ``linearization'', laying down the formalism for a 
similar discussion in the 4B case. The formal derivation is then presented 
in the Supplementary Information (SI). 
As such, what presented here can be considered as a short review of the 
results obtained for other MLPs, in particular for the powerspectrum case.
Crucially, we will maintain the equivalence with the ``internal coordinates
representation'' of the 3B term, Eq.~\eqref{eq:16}, so that one could freely 
move between the linear scaling formalism and the internal coordinates one, 
being the latter more advantageous for a small number of neighbours inside 
the cut-off volume. 

We start by remarking that the choice of Legendre polynomial as an expansion 
basis was primarily driven by their natural decomposition in terms of a sum 
of products of spherical harmonics, $Y_l^m$, namely
\begin{equation}\label{eq:17}
    P_l(\hat{\bm r}_1\cdot\hat{\bm r}_2) = \dfrac{4\pi}{2l+1} \sum_{m=-l}^{l}(-1)^m Y_l^m(\hat{\bm r}_1)Y_l^{-m}(\hat{\bm r}_2)\;.                                            
\end{equation}
By exploiting this property, and by combining Eq.~\eqref{eq:16} with 
Eq.~\eqref{eq:11}, one can prove that the 3B local energy term, 
$\varepsilon_i^{(3)}$, (defined so that $E_3= \sum_i \varepsilon_i^{(3)}$), 
can be written as
\begin{eqnarray}\label{eq:18}
    \varepsilon_i^{(3)} = \sum_{\substack{Z_1Z_2\\Z_1\ge Z_2}} \sum_{n_1 n_2 l}^{\text{unique}} b_{n_1 n_2 l}^{Z_1 Z_2 Z_i}\bigg[C^{(3),Z_1 Z_2}_{i n_1 n_2 l} - S^{(3),Z_1 Z_2}_{i n_1 n_2}\bigg]\:,
\end{eqnarray}
where the first sum runs over the atomic species present in the system. 
The coefficients $b_{n_1 n_2 l}^{Z_1 Z_2 Z_i}$ are simply proportional to 
$a_{n_1 n_2 l}^{Z_1 Z_2 Z_i}$, as shown in the SI, so that the equivalence 
in going from Eqs.~\eqref{eq:11}-\eqref{eq:13} to Eq.~\eqref{eq:18}, is preserved.
We refer to the coefficient $C^{(3),Z_1 Z_2}_{i n_1 n_2 l}$ as the coupling 
term, which is obtained by including, on top of proper pairs of neighbor atoms,
also the degenerate terms in which the central atom is allowed to interact 
twice with the same atom in the environment, namely we accept the cases 
$(j,j)_i$ in the sum of Eq.~\eqref{eq:11}. These ``self-interacting'' terms, 
$S^{(3),Z_1 Z_2}_{i n_1 n_2}$, must be then removed, and so they are subtracted 
in Eq.~\eqref{eq:18}. 

Explicitly, the coupling and the self interacting term are written as
\begin{equation}\label{eq:19}
    C^{(3),Z_1 Z_2}_{i n_1 n_2 l} = \dfrac{4\pi}{2l+1} \sum_{m=-l}^l (-1)^m A^{Z_1}_{in_1 lm} A^{Z_2}_{i n_2 l -m}\:,
\end{equation}
and 
\begin{equation}\label{eq:20}
     S^{(3),Z_1 Z_2}_{in_1 n_2} = \delta_{Z_1 Z_2} \sum_{j\in Z_1} \overline{P}_{n_1 ji}^{(\alpha,\beta)}\overline{P}_{n_2 ji}^{(\alpha,\beta)}\:,
\end{equation}
where $\delta_{Z_1 Z_2}$ is the Kronecker-delta. Here, we have adopted a 
``species-wise'' atomic basis from the one defined for the ACE potential (see
Ref.~\cite{ACE}), namely
\begin{equation}\label{eq:21}
    A_{inlm}^{Z} = \sum_{j \in Z} \overline{P}_{nji}^{(\alpha,\beta)}Y_l^m(\hat{\bm r}_{ji})\:,
\end{equation}
where the radial basis has been specialized to the double-vanishing Jacobi 
polynomials. Also, we note that Eq.~\eqref{eq:19} is proportional to the 
powerspectrum components~\cite{Bartok}, or to the analogous
rotationally-invariant product $B_{in_1 n_2 l}^{(2)}$ introduced for the ACE 
potential. The crucial point here is that the coupling term in Eq.~\eqref{eq:19} 
is written over the species-wise atomic basis of Eq.~\eqref{eq:21}. Since the
$A_{inlm}^Z$ basis scales linearly with the number of neighbours of the $i$-th 
atom, then we can evaluate the coupling term with a linear cost. This, together
with the fact that also the self energy scales linearly with respect to the 
number of neighbours, makes the computational scaling of the entire local 
energy, Eq.~\eqref{eq:18}, linear in the numbers of neighbor atoms.

Incidentally, we note that we can write the product of the double-vanishing 
Jacobi polynomials in Eq.~\eqref{eq:20} in terms of a linear combination of 
double-vanishing Jacobi polynomials, namely there are coefficients 
$c_n^{n_1n_2}$, such that
\begin{equation}\label{eq:22}
    \overline{P}_{n_1 ji}^{(\alpha,\beta)}\overline{P}_{n_2 ji}^{(\alpha,\beta)} = \sum_{n=2}^{n_1+n_2} c_n^{n_1n_2} \overline{P}_{n ji}^{(\alpha,\beta)}\:.
\end{equation}
The coefficients $c_n^{n_1n_2}$ are usually calculated by numerical integration. 
This shows that the self-energy term can be re-casted as a linear combination of 
$A_{in00}^{Z}$ too, and that it is, as expected, an effective 2B contribution.
However, given the possible different cut-off radii of the 2B and 3B potentials 
and the relative different truncation, $n_\text{max}$, we will keep the body orders 
as formally separated as possible, and we will not absorb the self-interaction 
terms back into lower body orders \cite{OrganicACE}.

Let us now define a practical extension of the atomic basis $A_{inlm}^Z$, so 
to simplify the discussion for higher-order terms. We define the JL-atomic 
basis as
\begin{equation}\label{eq:23}
    (J_p L_q)^{i,Z}_{n_1\ldots n_p l_1 m_1 \ldots l_q m_q} = \sum_{j\in Z}\bigg[\prod_{r=1}^p \overline{P}_{n_r ji}^{(\alpha,\beta)}\bigg] \bigg[\prod_{s=1}^q Y_{l_s}^{m_s}(\hat{\bm r}_{ji})\bigg]\:.
\end{equation}
This also includes the atomic basis $A_{inlm}^Z$, since
\begin{equation}\label{eq:23b}
    (J_1L_1)^{i,Z}_{nlm} = A_{inlm}^Z\:.
\end{equation}
However, the definition in Eq.~\eqref{eq:23b} allows us to take more than 
one double-vanishing Jacobi and one Legendre polynomial at once.

By looking at Eq.~\eqref{eq:22} (the same property holds for the Legendre
polynomials) one could appreciate how all the components of Eq.~\eqref{eq:23} 
can be reduced to linear combinations of $A_{inlm}^Z$. Therefore, the definition 
of the JL-basis could appear unnecessary. However, since the coefficients
$c_n^{n_1n_2}$ must be evaluated by integration, it can be more convenient 
to use directly the JL-atomic basis instead than performing the necessary 
integrations and contractions. It is important to notice that evaluating the
elements of the JL-atomic basis is still linear with the number of neighbours 
of the $i$-th atom: the only scaling affected is in terms of the number of 
the components involved, namely the number of polynomials in the product.

Finally, we can now write the coupling term and the self energy over the 
JL-atomic basis as
\begin{equation}
\begin{dcases}
C^{(3),Z_1 Z_2}_{i n_1 n_2 l} = \dfrac{4\pi}{2l+1} \sum_{m=-l}^l (-1)^m (J_1L_1)^{i,Z_1}_{\substack{n_1 l m}} (J_1L_1)^{i,Z_2}_{\substack{n_2 l-m}}\:,\\
S^{(3),Z_1 Z_2}_{in_1 n_2} = \delta_{Z_1 Z_2} (J_2L_0)^{i,Z_1}_{\substack{n_1 n_2}}\:.
\end{dcases}
\end{equation}
The JL-atomic basis will be used as the general framework for the analogous 
analysis of the linear scaling in the 4B case.

\subsection{The Four-Body Term}\label{sec:fourb}

For the 4B case we will follow the very same steps presented for the 3B one. 
We start by expanding the 4B energy contribution, $E_4$, as
\begin{equation}\label{eq:25}
    E_4 = \sum_{i}^\text{atoms} \sum_{(j,k,p)_i}v^{(4)}_{jkpi}(r_{ji},r_{ki},r_{pi},s_{jki},s_{kpi},s_{jpi})\:,
\end{equation}
where, analogously to $E_3$ in Eq.~\eqref{eq:11}, the second sum runs over 
all the triplets of atoms in the neighborhood of the $i$-th atom. As for the 
3B case, $v^{(4)}_{jkpi}$ is a shorthand form for $v^{(4)}_{Z_j Z_k Z_p Z_i}$.

The 4B potential, $v^{(4)}$, depends on 3 distance and 3 angles, so that 
any triplets of atoms in the neighborhood of the $i$-th one is uniquely 
determined up to a reflection. The JL-4B expansion is then simply obtained 
by generalizing Eq.~\eqref{eq:16} to the case in which we have three 
double-vanishing Jacobi polynomials and as many Legendre polynomials, 
namely
\begin{eqnarray}\label{eq:26}
    &&v^{(4)}_{jkpi}(r_{ji},r_{ki},r_{pi},s_{jki},s_{kpi},s_{jpi}) = \\
    &&= \sum_{\substack{n_1 n_2 n_3\\l_1 l_2 l_3}}^{\text{unique}} a_{\substack{n_1 n_2 n_3\\l_1 l_2 l_3 }}^{jkpi} \sum_{\text{symm.}}\bigg({\overline{P}}_{n_1 ji}^{(\alpha,\beta)}\overline{P}_{n_2ki}^{(\alpha,\beta)}\overline{P}_{n_3pi}^{(\alpha,\beta)}P_{l_1}^{jki}P_{l_2}^{jpi}P_{l_3}^{kpi}\bigg)\nonumber\:.
\end{eqnarray}
The range of the Jacobi indexes is again $[2, n_\text{max}]$, while that of the
Legendre ones is $[0,l_\text{max}]$, where both $n_\text{max}$ and 
$l_\text{max}$ require optimization. By adopting the same formalism of 
Eq.~\eqref{eq:16}, the symmetries of the potential are implemented in the 
expansion by construction. As for the 3B case, we have that
$a_{\substack{n_1 n_2 n_3\\l_1 l_2 l_3 }}^{jkpi}$ is a shorthand for $a_{\substack{n_1 n_2 n_3\\l_1 l_2 l_3 }}^{Z_jZ_kZ_pZ_i}$. 

It is useful to explicitly investigate the symmetries for the case in 
which the atoms in the neighborhood belong to the same species. By associating 
the Jacobi indexes $n_1$, $n_2$ and $n_3$ to the first, second and third distances
respectively, and analogously associating the Legendre indexes to the scalar
products, we impose the following symmetries on the expansion coefficients
\begin{eqnarray}\label{eq:27}
    &&a_{\substack{n_1 n_2 n_3\\l_1 l_2 l_3}} = a_{\substack{n_2 n_1 n_3\\l_1 l_3 l_2}} = a_{\substack{n_3 n_2 n_1\\l_3 l_2 l_1}} = \nonumber\\
    &&=a_{\substack{n_1 n_3 n_2\\l_2 l_1 l_3}}= a_{\substack{n_2 n_3 n_1\\l_3 l_1 l_2}} = a_{\substack{n_3 n_1 n_2\\l_2 l_3 l_1}}\:.
\end{eqnarray}
The first identity states that, when exchanging the first two atoms, we have 
to simultaneously exchange the relative distances from the central atom (swapping
the Jacobi indexes $n_1$ and $n_2$) and the angles formed with the remaining atom
(exchanging the Legendre indexes $l_2$ and $l_3$). All the other identities can 
be interpreted in a similar way. The equivalences in Eq.~\eqref{eq:27} give us 
the unique set of indexes to use in Eq.~\eqref{eq:26}, so that the number of
parameters to learn is reduced by roughly a fact six. The second sum in
Eq.~\eqref{eq:26} will then run over all the indexes permutations involved 
in Eq.~\eqref{eq:27}, mostly resulting in a sum of six terms, similarly to what 
was explicitly shown in Eq.~\eqref{eq:15}.

We conclude this section by remarking that the use of double-vanishing 
polynomials in the 4B-JL expansion allows us to implement an even more severe
reduction in the number of free coefficients compared to the 3B case.

\subsection{4B Linear Scaling: connection with the Bispectrum}

A linear scaling with the number of atoms in the neighbourhood volume can 
also be achieved for the 4B case. Indeed, this is even more important than
for lower-body orders, since otherwise the scaling would be cubic with the 
number of neighbors. The backbone of the demonstration is similar to the 
one adopted for the 3B case, so that, by using the property of Eq.~\eqref{eq:16} 
and the JL-atomic basis defined in Eq.~\eqref{eq:23}, we can write the 
local energy term, $\varepsilon^{(4)}_i$, as
\begin{eqnarray}\label{eq:28}
    &&\varepsilon_i^{(4)} = \sum_{Z_1\ge Z_2\ge Z_3} \sum_{\substack{n_1 n_2 n_3\\ l_1 l_2 l_3}}^{\text{unique}} b_{\substack{n_1 n_2 n_3\\ l_1l_2l_3}}^{Z_1 Z_2 Z_3 Z_i}\times\\
    &&\qquad \qquad\qquad\qquad\times \Bigg[C^{(4),Z_1 Z_2 Z_3}_{i ,\substack{n_1 n_2 n_3\\ l_1l_2l_3}} - S^{(4),Z_1 Z_2 Z_3}_{i ,\substack{n_1 n_2 n_3\\ l_1l_2l_3}}\Bigg]\:,\nonumber
\end{eqnarray}
where the coupling term for the 4B is given by
\begin{eqnarray}\label{eq:29}
    &&C^{(4),Z_1 Z_2 Z_3}_{i ,\substack{n_1 n_2 n_3\\ l_1l_2l_3}} = \dfrac{(4\pi)^3}{(2l_1+1)(2l_2+1)(2l_3+1)}\times\\
    &&\qquad\times\sum_{m_1 m_2 m_3}(-1)^{m_1 + m_2 + m_3}(J_1L_2)^{i,Z_1}_{\substack{n_1 l_1 m_1 l_2 -m_2 }}\times\nonumber\\
    &&\qquad\qquad\qquad\times(J_1L_2)^{i,Z_2}_{\substack{n_2 l_3 m_3 l_1 -m_1 }}(J_1L_2)^{i,Z_3}_{\substack{n_3 l_2 m_2 l_3 -m_3 }}\:. \nonumber
\end{eqnarray}
The corresponding expression for the self-energy, $S^{(4)}_i$, is more involved 
and, for the sake of brevity, is reported in the SI. Here, we just wish to mention 
that it is obtained from linear combinations of products of the basis terms
$(J_1L_2)$, $(J_2L_2)$ and $(J_3L_0)$.

The coupling scheme described in Eq.~\eqref{eq:29} differs from the 
bispectrum-components coupling scheme \cite{Bartok,SNAP}, while being strictly
related to it. Indeed, the bispectrum writes in the ACE flavour \cite{ACE} as,
\begin{eqnarray}
    &&B_{i\substack{n_1n_2n_3\\l_1 l_2 l_3}}^{(3),Z_1 Z_2 Z_3} =\sum_{m_1 m_2 m_3}\mqty(l_1 & l_2 & l_3\\m_1 &m_2& m_3)\times\nonumber\\
    &&\qquad\qquad\qquad \times A^{Z_1}_{in_1 l_1 m_1}A^{Z_2}_{in_2 l_2 m_2}A^{Z_3}_{in_3 l_3 m_3}\:,
\end{eqnarray}
where the 3j-Wigner symbol \cite{Angular} is introduced and $A_{inlm}$ is the
atomic basis of Eq.~\eqref{eq:21}. Furthermore, the JL-atomic basis terms,
$(J_1L_2)$, can be written as a linear combination of the $A^Z_{inlm}$,
\begin{eqnarray}\label{eq:31}
    &&(J_1L_2)^{i,Z}_{n l_1 m_1 l_2 m_2 } = \nonumber\\
    &&\quad= \sum_{lm} (-1)^m \sqrt{\dfrac{(2l+1)(2l_1+1)(2l_2+1)}{4\pi}}\times\\
    &&\qquad\qquad \times\mqty(l_1 &l_2 &l\\0 &0& 0)\mqty(l_1 &l_2 &l\\m_1 &m_2& -m)  A^Z_{inlm}\:.\nonumber
\end{eqnarray}
This is directly derived from the product rule for two spherical harmonics 
(see SI). From this expression, one could write the coupling terms $C^{(4)}_i$ 
as a linear combination of bispectrum components $B_i^{(3)}$, as reported in 
the SI. This linear combination represents the way of combining the bispectrum
components so that the final result is explicitly written in terms of internal
coordinates only. Crucially, our argument shows that adopting the coupling scheme
in $C^{(4)}_i$ could give an advantage over the bispectrum components, since it
allows us to maintain the equivalence between the expression in Eq.~\eqref{eq:29} 
and the analogous one from Eqs.~\eqref{eq:25}-\eqref{eq:26}. Then, an intuitive 
and equivalent closed expression (in terms of internal coordinates) remains 
available for any case in which the number of atoms in the neighbours is 
relatively small, so that one could opt between Eq.~\eqref{eq:28} and 
Eqs.~\eqref{eq:25}-\eqref{eq:26} at need.

\subsection{The Five-Body Term}
The 5B term can be obtained by direct generalization of the 4B case. Indeed, 
the procedure is analogous, namely the energy contribution, $E_5$, is partitioned 
in local components, which consist of a sum of local 5B potentials, $v^{(5)}_{jkpqi}$. 
These depend on four distances and six angles, so that they can be expanded as 
a linear combination of products of four double-vanishing Jacobi polynomials 
and six Legendre polynomials. The resulting expression is analogous to the one
obtained in Eq.~\eqref{eq:18} for the 4B case. The symmetry properties of the
potentials are also treated in the same way, resulting in a reduction of the 
number of coefficients up to roughly a factor 24 when dealing with identical 
atoms.

\subsection{Behaviour at the origin}\label{sec:origin}
A common practice in MLPs is to introduce an external function so to 
impose a repulsive behaviour when the interatomic distance becomes small. 
Here, since we are using the double-vanishing Jacobi polynomials for all 
body orders beyond two, the only term affecting the behaviour at small 
distances is the two-body one, given in Eq.~\eqref{eq:5}. We can then 
obtain some insight into the behaviour of the potential by evaluating 
Eq.~\eqref{eq:5} at the origin. Indeed, if $r_\text{min} = 0$, we 
obtain
\begin{equation}
    v^{(2)}_{Z_j Z_i}(0) =  \sum_{n=1}^{n_\text{max}}a_n^{Z_j Z_i} \widetilde{P}_n^{(\alpha,\beta)}(1)\:.
\end{equation}
From the identity
\begin{equation}
    \widetilde{P}^{(\alpha,\beta)}_n(1) = \binom{n+\alpha}{n} - (-1)^n\binom{n+\beta}{n}\:, \nonumber   
\end{equation}
we can conclude that the magnitude of the potential at the origin can become 
very large for an high enough $n$. Therefore, by biasing the hyper-parameters 
so that the potential is positive at the origin, we can produce a strongly
repulsive behaviour almost by construction, with no use of any external function. 

This observation must be checked on a case-by-case base, an operation that can 
be performed visually by simply looking at the potential. Indeed, once the best
expansion coefficients are available, it is possible to plot the function
\begin{equation}\label{eq:33}
    v^{(2)}_{Z_j Z_i}(x) =  \sum_{n=1}^{n_\text{max}}a_n^{Z_j Z_i} \widetilde{P}_n^{(\alpha,\beta)}\left(\cos\left(\pi x/r_\text{cut}\right)\right)\:,
\end{equation}
and analyse the behaviour near the origin. Since small distances are usually in 
an extrapolation region of the potential, with little to no data corresponding to 
such distances present in the training set, a visual investigation of the 2B 
potential could also return us some intuition on the behaviour of the model when 
dealing with extrapolation attempts to unseen atomic distributions.

\subsection{Forces and Stress}
In this section we outline the general recipe to calculate the forces and the
virial-stress tensor. Given the linearity of the expressions associated with 
the JL expansion, one only needs the derivative of the (double-)vanishing 
Jacobi and of the Legendre polynomials, from which all the relevant quantities 
can be evaluated.

Since the multi-body expansion of the energy, Eq.~\eqref{eq:1}, and the fact that
$E_1$ is just an energy offset, the $n$-body contribution to the force of an atom at position ${\bm r_a}$, is given 
by
\begin{equation}
    \bm F_a^{(n)} = -\pdv{E_n}{\bm r_a}\:,
\end{equation}
whereas the total force is obtained by summing up over all the $n$-body 
contributions, $\bm F_a = \sum_n F_a^{(n)}$. 

As it can be seen from Eq.~\eqref{eq:4} and Eq.~\eqref{eq:5}, the evaluation of 
the 2B force contribution, $F_a^{(2)}$, requires only the application of the 
chain rule and the derivative of the vanishing polynomials, namely
\begin{eqnarray}\label{eq:35}
    &&\dv{x} \widetilde{P}_n^{(\alpha,\beta)}(\cos(x)) = \dv{x} P_n^{(\alpha,\beta)}(\cos(x))=\\
    &&= -\dfrac{\alpha+\beta+n+1}{2}\sin(x) P^{(\alpha+1,\beta+1)}_{n-1}(\cos(x))\:\nonumber.
\end{eqnarray}
This expression shows that the derivative of the potential smoothly vanishes 
($\bm F_a^{(2)}=0$) at the cut-off radius and at the origin (when $x = 0,\pi$).
Furthermore, from Eq.~\eqref{eq:35} we can appreciate that the force can be 
written solely in terms of Jacobi polynomials. This results in a linear expansion 
that can be easily implemented or analytically investigated.

Analogously, we can evaluate the 3B contribution to the forces by differentiating
the $E_3$ term. This implies that we need to calculate [see Eq.~\eqref{eq:13} 
and Eq.~\eqref{eq:16}]
\begin{eqnarray}
    &&\pdv{}{\bm r_a}\sum_{\text{symm.}}\bigg({\overline{P}}_{n_1 ji}^{(\alpha,\beta)}\overline{P}_{n_2ki}^{(\alpha,\beta)}P_l^{jki}\bigg) = \nonumber\\
    &&= \sum_{\text{symm.}}\pdv{}{\bm r_a}\bigg({\overline{P}}_{n_1 ji}^{(\alpha,\beta)}\overline{P}_{n_2ki}^{(\alpha,\beta)}P_l^{jki}\bigg)\:,
\end{eqnarray}
where we are able to exchange the sum and the derivative, since the former 
acts only on the Jacobi and Legendre indexes. Therefore, in evaluating the
derivative of the product, we can use again the chain-rule and the differentiation
formula for the Legendre polynomials, namely
\begin{equation}
    \dv{x}P_l(x) = \dv{x} P^{(0,0)}_l(x) = \dfrac{l+1}{2}P^{(1,1)}_{l-1}(x)\:,
\end{equation}
where we have used the fact that the Legendre polynomials are obtained from the 
Jacobi polynomials by setting $\alpha=\beta = 0$. Finally, we also need the
differentiation rule for double-vanishing Jacobi polynomials
\begin{eqnarray}
     &&\dv{x} \overline{P}_n^{(\alpha,\beta)}(\cos(x))=\nonumber\\
     &&= -\dfrac{\sin(x)}{2}\Bigg((\alpha+\beta+n+1) P_{n-1}^{(\alpha+1,\beta+1)}(\cos(x))+\nonumber\\
     &&\qquad\qquad\qquad-(\alpha+\beta+2)\dfrac{\widetilde{P}_{n}^{(\alpha,\beta)}(-1)}{\widetilde{P}_{1}^{(\alpha,\beta)}(-1)}\Bigg)\:.
\end{eqnarray}

The 4B and 5B contributions to the forces are evaluated in the same way, and these
do not introduce any further ingredient to obtain an analytical form. The 
expression for the forces in the case of the JL-atomic basis will be explicitly
discussed in future works. Finally, we can also obtain the virial-stress tensor 
by mean of the formula discussed in reference~\cite{thompsonStress} 
[see Eq.~(25)].

\subsection{Linear Regression}
In order to select the optimal expansion coefficients for each body term, we minimise the widely used loss function
\begin{eqnarray}
    &&L = \norm{\bm E -  \bm J_E\bm a}^2_2+ c_F \norm{\bm F -  \bm J_F\bm a}^2_2+ c_W \norm{\bm W -  \bm J_W\bm a}^2_2 \:\nonumber\:,
\end{eqnarray}
where the vector $\bm E$ represents all the energies in the training set 
(obtained by \emph{ab-initio} calculations), $\bm a$ is the vector of all 
the coefficients of the expansion, $\bm J_E$ is the matrix, whose rows contain 
the set of descriptors for one configuration of the training set. Similarly 
$\bm F$ is the vector of all the forces of the dataset, while $\bm J_F$ are 
the appropriate differentiated descriptors. Note that, explicitly, we will train 
on each components of the forces for each atom in the system. This means that, 
if the $i$-th configuration has $N_i$ atoms, we will have $3N_i$ forces associated
to that configuration. The vector of the components of the stress tensor for 
each training point is $\bm W$. In this case, we will train independently on 
each of the six components, for any of the configurations in the training 
set. Finally $c_F$ and $c_W$ are coupling constants to be optimized, and 
$\norm{\cdot}^2_2$ is the square of the vector 2-norm. While the use of a 
multi-target scheme, embedded in a non-linear function, can be used to increase 
the accuracy of the model \cite{parameterACE}, we remark that we follow here 
a simple linear approach.

The minimization procedure that will be adopted for the remainder of this 
work, where results on a mono-species system are shown, will be based on the
Singular Value Decomposition (SVD). We stress that we will not regularise the
energy offset, $E_1$. Furthermore, instead of using the total energies, we will
always consider the energy per atom in the training set. 

Here we wish to remark that the coupling constants, $c_F$ and $c_W$, can also
depend on the specific configuration, a fact that can be seen as a 
configuration-wise re-scaling of the descriptors and targets. This is useful, 
in particular, when the configurations have a different number of atoms. As 
a direct example, the loss function used in the next section, will have all 
the forces and the relative descriptors divided by $\sqrt{3N_i}$, where $N_i$ 
is the number of atoms in the configuration. This is performed in order to 
weight the energies, forces and stress, on a similar footing in the minimization
procedure. Another advantage of such a configuration-wise weighting scheme is 
that the energy offset per atom, $E_1$, can be written analytically in terms of 
the per-atom average energy, average descriptors and the linear fitting 
coefficients.

\begin{figure*}
\centering
\includegraphics[width=\textwidth]{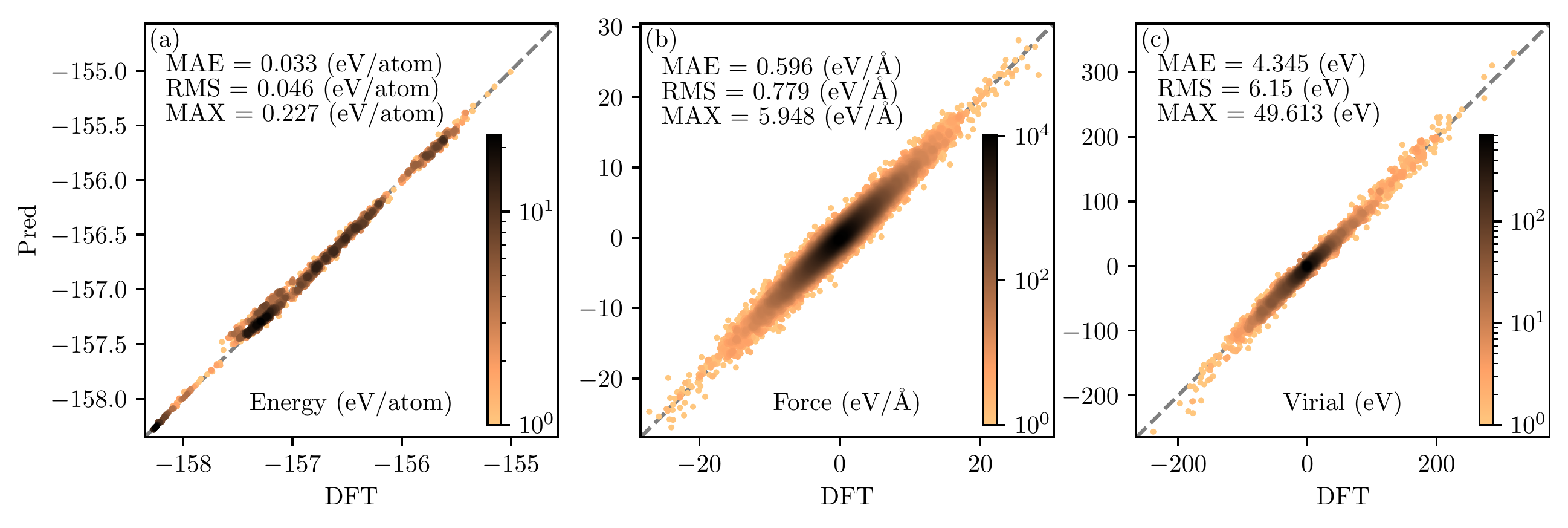}
\caption{Parity plots computed over the test set for the (a) energies, 
(b) forces, (c) virial stress. The Mean Absolute Errors (MAEs) and Root
Mean Square Error (RMSE) are reported for each plot, alongside with the error 
on the worst prediction. The color code indicates the data density (number of 
points).}
\label{fig:5}
\end{figure*}

\begin{table}[t]\centering
\ra{1.2}
\setlength{\tabcolsep}{6pt} % Default value: 6pt
\begin{tabular}{@{\extracolsep{3pt}}lccc@{}}
\hline \hline
\T \B   & \multicolumn{1}{c}{Two Body} &   \multicolumn{1}{c}{Three Body} & \multicolumn{1}{c}{Four Body}\\
\cline{2-2} \cline{3-3} \cline{4-4}

$ n_\text{max} $ \T \B & 10&6 &4\\
        $ l_\text{max} $ \T \B & -- & 5 &3 \\
        $ r_\text{cut}\, ($\AA$)$ \T \B &3.7&3.7 &3.7 \\
        $\alpha = \beta$ \T \B &  1&1&1\\
        \# of features  & 10 & 90 &364\\

\cline{1-4}
\end{tabular}
\caption{Details of the JLP trained on the carbon dataset 
from Ref.~\cite{gapC}. In order to reduce the number of hyper parameters, we 
fix $\alpha$ and $\beta$ to be equal, and $r_\text{min}=0$. The model is relatively
compact and comprises 465 (464 plus the intercept) features.\label{tab:HP_C}}
\end{table}

\section{A JLP for Carbon}

As an application of the method described here, we have fitted a JLP on the 
carbon dataset used to fit the GAP17 potential of reference~\cite{gapC}. We 
have opted for this dataset, since it presents several challenges. Firstly, 
the dataset is made of different phases of carbon, ranging from crystalline 
structures (graphene, graphite, diamond), to surfaces and amorphous phases. 
In addition, some phases present a relative large distance for the decay of 
the forces between two atoms, as explicitly shown in the same Ref.~\cite{gapC}. 
This is mirrored in the choice of the appropriate cut-off radius. For the fit 
we have removed all the carbon dimers (config\_type=cluster) and any structures
with absolute maximum force components greater than 30 eV/\AA. In total we 
have thus removed 37 structures of which 30 are the carbon dimers used to fit 
the two body GAP and 7 other structures, which do not satisfy the maximum force 
criteria. The remaining 4,043 structures are split into a training set of 2,830 
and a testing one of 1,213. The structure index of all the training and testing
structures are given in the SI.

We use energy, forces and virial stress to fit the linear model. The 
hyper-parameters for the final potential are summarised in Table~\ref{tab:HP_C}.
Following the analysis on the locality of Ref.~\cite{gapC}, we have kept the same
cut-off radius as for the GAP17 model, namely 3.7~\AA. The coupling constant 
$c_F$ and $c_W$, of the loss function, are $0.5$ and $0.075$ respectively. 
Finally, the descriptors have been calculated in their internal coordinate form 
and the cluster expansion is truncated at the four-body order. This gives us
a potential defined over 465 features.
\begin{figure}
    \centering
    \includegraphics[width=0.9\columnwidth]{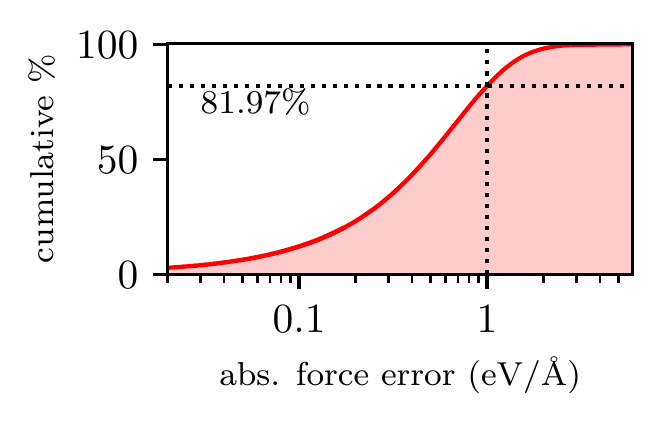}
    \caption{The cumulative distribution of the test-set predicted forces. 
    The model shows that approximately 81.97\% of the structures have an error
    below 1 eV/\AA.}
    \label{fig:cumerr}
\end{figure}

For the fitted model, we find that the training-set Root Mean Squared Errors
(RMSEs) are 43.9~meV/atom for the energy, 0.779~eV/\AA\ for the forces 
and 6.62~eV for the stress. As shown in Fig.~\ref{fig:5}, reporting the parity 
plots for the test set, the corresponding RMSEs are 46.6~meV/atom for the 
energy, 0.781~eV/\AA\ for the forces and 6.15~eV for the stress, namely they 
are of the same quality as for the training set (the parity plots for the 
training set are reported in the SI). We observe that the structures, which 
deviate the most from the energy-parity plot in Fig.~\ref{fig:5}(a), correspond 
to all carbon in the amorphous phase. These appear to be slightly more difficult 
to be dealt with by the JLP. Furthermore, we wish to remark that, as it can be
appreciated in Fig.~\ref{fig:5}(c), the predicted components of the virial-stress 
appear to be in good agreement with the DFT ones.

In Fig.~\ref{fig:cumerr} we report the cumulative distribution of the the error 
on the forces for the test set. The curve represents the percentage of structures,
which have an error below the one indicated. As a reference, we explicitly consider
the case of 1 eV/\AA, which was taken as reference for the GAP17 potential (see
Ref.~\cite{gapC}). The remarkable high value of 81.97\% shows the capability of 
the JLP in correctly predicting the forces components.
\begin{figure}
    \centering
    \includegraphics[width=0.9\columnwidth]{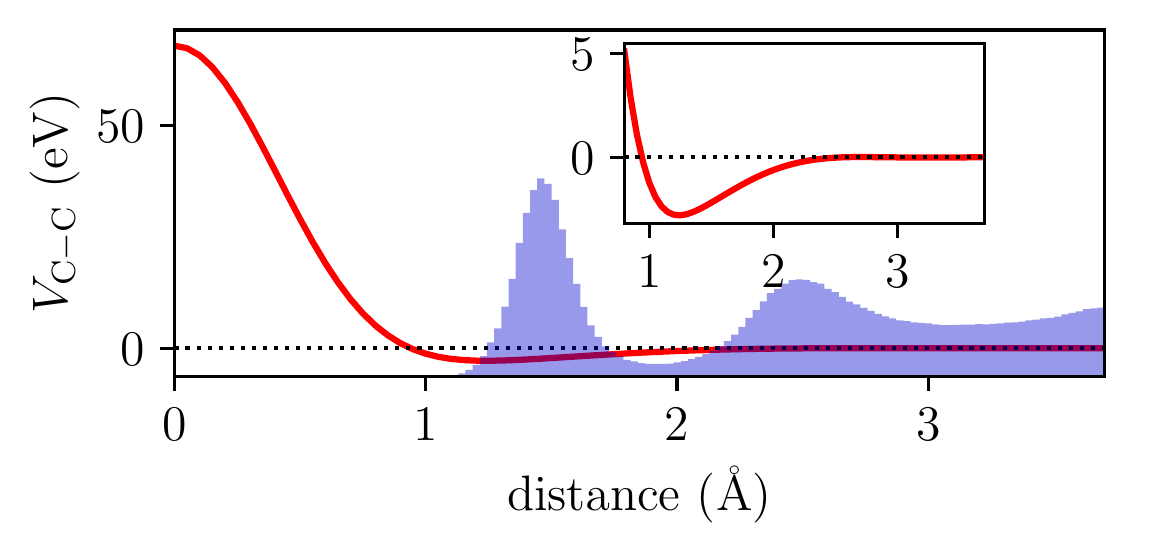}
    \caption{Reconstruction of the 2B potential from Eq.~\eqref{eq:33} (red curve).
    The insert show a magnification around the minimum, while the histogram 
    reports the pair-distance distribution of the entire dataset. Qualitatively,
    the potential shows a strong repulsive behaviour for small distances and a minimum, which is consistent with the position of the first peak in the 
    pair-distances distribution.}
    \label{fig:recon}
\end{figure}

As remarked in Section~\ref{sec:origin}, the JL potential naturally shows a
repulsive behaviour at short distance, without the inclusion of any external 
fast-varying function. This is made clear in Fig.~\ref{fig:recon},
where we show the C-C potential obtained by plotting Eq.~\eqref{eq:33} with the
fitted 2B-expansion coefficients. From the figure one can clearly identify the
short-distance repulsive behaviour of the 2B potential, which arises naturally 
from the 2B expansion coefficients. Furthermore, the potential shows a shallow
minimum close to the the first peak in the radial distribution function computed 
over the fitting dataset (blue shadow). We stress here once again that the 
repulsive behaviour is completely determined by the 2B coefficients, since all 
the other body-order terms are written in term of the double-vanishing Jacobi 
polynomials, which vanish at short distances. As a consequence, 
Fig.~\ref{fig:recon} gives us complete information about the repulsive behaviour 
of the entire potential.

\begin{figure}
    \centering
    \includegraphics[width=0.9\columnwidth]{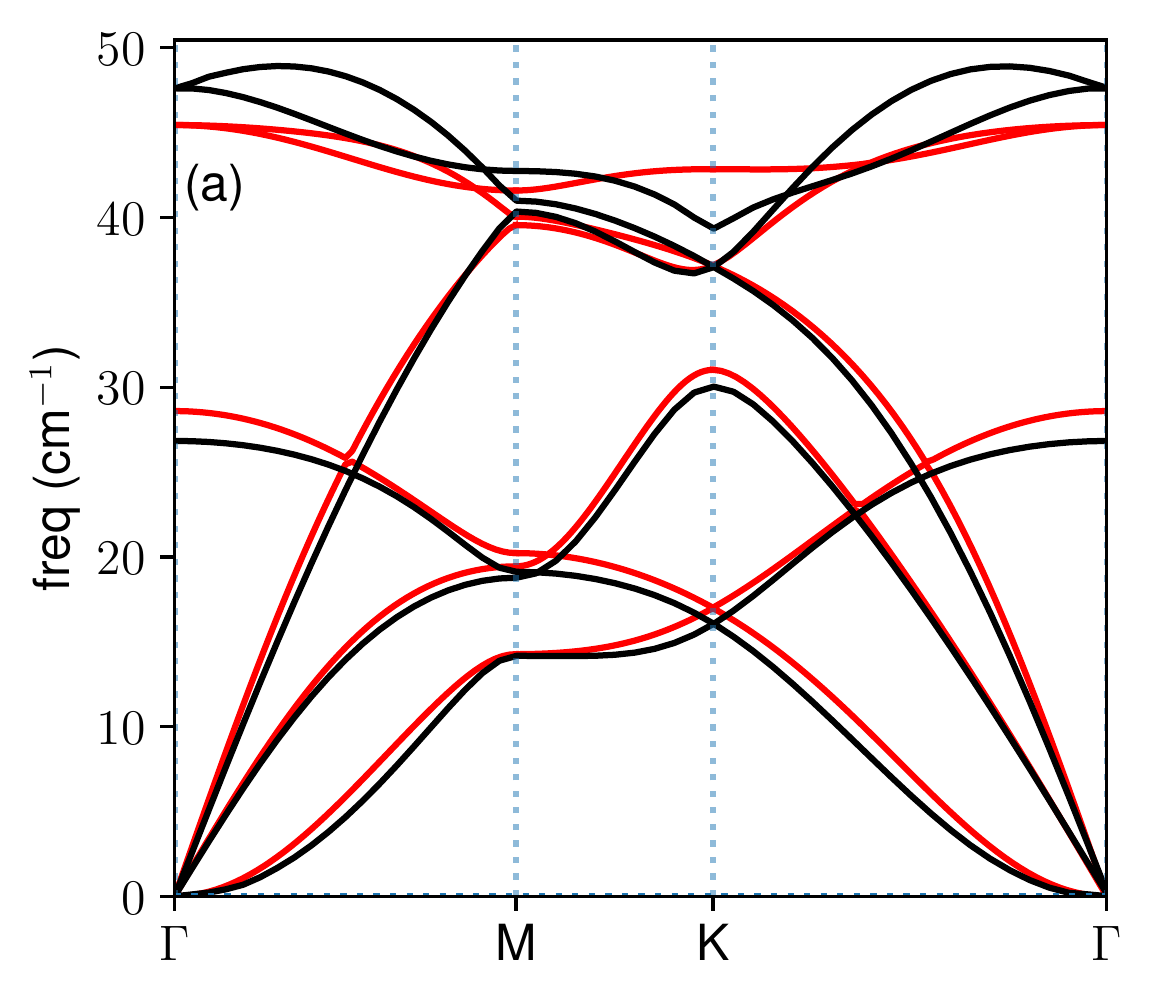}
    \includegraphics[width=0.9\columnwidth]{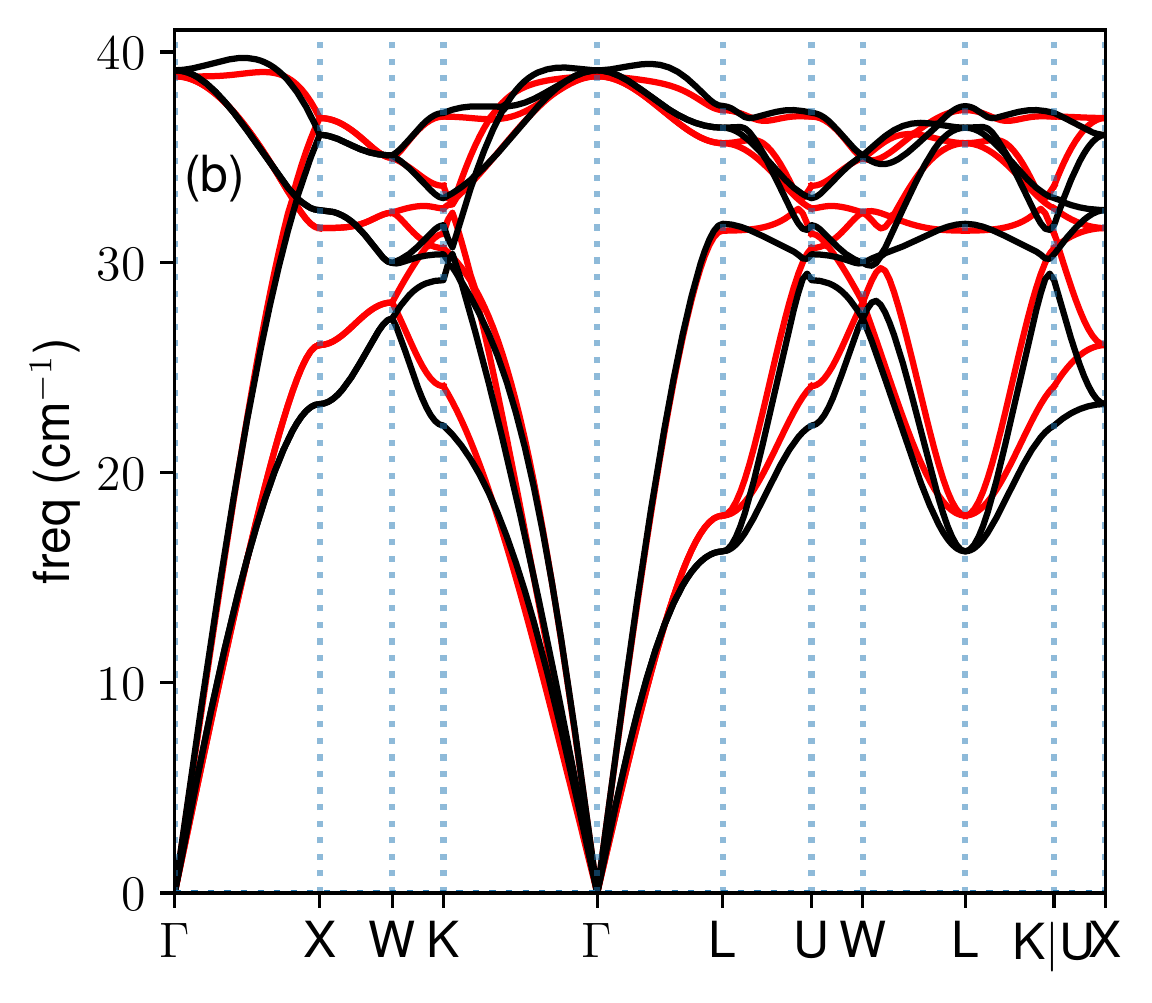}
    \caption{Phonon spectra for (a) graphene and (b) diamond computed with the
    optimised JLP described in the text (red lines). The reference DFT calculations 
    (black lines) have been obtained with density functional perturbation theory as 
    implemented in the {\sc abinit} code.}
    \label{fig:phonons}
\end{figure}

We then employed the optimized JLP, to predict the phonons dispersion curves for
graphene and diamond, using the phono3py package \cite{phono3py1,phono3py2}. The 
results are reported in Fig.~\ref{fig:phonons}, where the reference phonon 
dispersion for crystalline diamond (mp-66) was obtained from materials project 
\cite{materialsproject} and for graphene was obtained from the phonon website 
\cite{phononwebsite}. These reference calculations have performed using density
functional perturbation theory using the {\sc abinit} code \cite{abinitdfpt}. 

%\sout{obtained by DFT, are computed using Vienna Ab initio Simulation Package
% (VASP, version-5.4.4) \cite{vasp1,vasp2}. For this, we have adopted the LDA
% pseudopotentials (version 54, PAW C 22Mar2012), the energy cutoff is set to 
% \num{650}~eV, while the $k$-grid spacing is \num{0.1} \AA$^{-1}$. The electronic
% convergence criteria threshold is $10^{-8}$~eV. Before performing the phonon 
% calculation, structural relaxation was performed until the maximum force component 
% on each atom was below \num{0.001} eV/\AA. During the relaxation all the unit cell parameters and the atomic positions were allowed to move.}

As one can appreciate from the figure, the agreement between the JLP-computed 
phonon bands and the DFT reference ones is quite remarkable, for both the acoustic 
and optical branches. The largest disagreement is generally found for the optical
branches and it is of the order of 2~cm$^{-1}$ (see, for instance, the graphene bands 
at around 45~cm$^{-1}$). Note that this is a particular challenging test, since
the training dataset has an energy spread of several eV/atom, while the energy
differences computed in the finite-difference scheme used here are a few meV/atom
from the equilibrium energy. This means that our JLP is able to describe, on the same
footing, both the low-energy physics of crystalline carbon around equilibrium, and
high-energy liquid and amorphous structures. Note also that perfect agreement is 
not even expected. In fact, the DFT dataset used to train the JPL model was 
obtained with the {\sc castep} code~\cite{castep} and the phonon via finite differences,
while our DFT reference has been generated with {\sc abinit}~\cite{abinitdfpt} and 
density functional perturbation theory. Additional differences can also be 
ascribed to the different pseudopotentials used and to details in the DFT 
implementation.

\section{Conclusions}

In conclusion, we have introduced all the necessary formalism to develop a general
cluster expansion for the total energy, where the different body-order terms are
systematically separated. This is designed for the short-range chemical-bond-related
part of the total energy, which is written as the sum of individual atomic 
contributions. The core idea is then that of expanding the different body-order 
terms, representing the inter-atomic distances over Jacobi polynomials and the
structural angles over Legendre polynomials, which are a special case of the 
Jacobi ones. This is an extremely general representation, giving us ample
flexibility when constructing the potential. 

An important feature is that one can impose both constrains and symmetries on 
the coefficients of the expansion, a practice that allows us to implement desired
behaviours of the potential in a natural way. For example, one can impose the
potential to vanish at a desired cut-off distance, by simply imposing a set
of conditions over the zero-order coefficients of the Jacobi polynomial expansion.
In a similar way, one can constrain the expansion of all the body-order
terms larger than two to vanish at the origin, so that the short-distance 
behaviour of the potential is solely determined by the two-body contribution. 
This, in turn, can be designed to display a repulsive behaviour at short 
distances. 

Furthermore, the implementation of physical symmetries over the
expansion allows us to drastically reduce the number of independent
coefficients to determine. As a result, the number of features that uniquely 
define the cluster expansion is demonstrated to scale linearly with the number
of atoms in the cut-off volume. The demonstration of such scaling is rooted in
the decomposition of the Legendre polynomials over spherical harmonics,
a feature, which allows us to map our representation on known many-body 
atomic bases such as the powerspectrum, the bispectrum and those introduced
in the atomic cluster expansion method.

The formalism introduced here is put to the test for a quite complex dataset,
namely the carbon one used to construct the GAP17 potential. This comprises
crystalline graphite and diamond, as well as a multitude of liquid and amorphous
carbon structures. We then show that a four-body relatively compact model, 
containing 465 features and trained over energies, forces and stress tensor, 
is capable of achieving extremely competitive RMSEs across all quantities. 
Furthermore, the same potential reproduces quite accurately the zero-temperature
phonon band structure of both graphene and diamond, demonstrating accuracy both
at low and high energy. We believe that the JLP introduced here adds to the 
burgeoning field of machine-learning potentials, bringing a versatile tool where
symmetry and constraints can be implemented in a natural and efficient way. The
ability to separate the different body orders and the possibility to construct
relatively compact models, make the JLP a strong candidate for the calculation 
of potential energy surfaces both in data-reach and data-poor situations.

\begin{acknowledgments}
This work has been supported by the Irish Research
Council Advanced Laureate Award (IRCLA/2019/127),
and by the Irish Research Council postgraduate program
(MC). UP acknowledge the Qatar National
Research fund for additional financial support (grant
no. NPRP12S-0209-190063). We acknowledge the DJEI/DES/SFI/HEA Irish
Centre for High-End Computing (ICHEC) and Trinity
Centre for High Performance Computing (TCHPC) for
the provision of computational resources. We acknowledge support 
from ICHEC via the academic flagship program (Project Number - EuroCC-AF-3).
We have extensively used numpy~\cite{numpy}, cython~\cite{cython}, scikit-learn~\cite{sklearn} and matplotlib~\cite{matplotlib}.
\end{acknowledgments}

\appendix
\section{Proof of property \eqref{eq:10}}\label{app:A}
We derive here a series expansion for the vanishing Jacobi polynomials and we will prove the property \eqref{eq:10}. The series expansion for the Jacobi Polynomials is (Ref. \cite{Abramovitz})
\begin{equation}
    P_n^{(\alpha,\beta)}(x) = \dfrac{1}{2^n}\sum_{j=0}^n \binom{n+\alpha}{j}\binom{n+\beta}{n-j}(x-1)^{n-j}(x+1)^j\:.
\end{equation}
Performing the substitution $x\rightarrow \cos x$, where $x = \pi (r-r_\text{min})/(r_\text{rcut}-r_\text{min})$, we get
\begin{eqnarray}
    &&P_n^{(\alpha,\beta)}(\cos x) \\
    &&= \sum_{j=0}^n (-1)^{n-j}\binom{n+\alpha}{j}\binom{n+\beta}{n-j}\sin^{2(n-j)}(x/2)\cos^{2j}(x/2)\nonumber\: .
\end{eqnarray}
Evaluating the expression on $x=\pi$, which means to evaluate the polynomial at the cut-off, makes all the terms of the summation vanish expect for the case $j=0$. Therefore, by mean of \eqref{eq:6}, we get a series expansion for the vanishing Jacobi Polynomials.
\begin{eqnarray}
    &&\widetilde{P}_n^{(\alpha,\beta)}(\cos x) \\
    &&= \sum_{j=1}^n (-1)^{n-j}\binom{n+\alpha}{j}\binom{n+\beta}{n-j}\sin^{2(n-j)}(x/2)\cos^{2j}(x/2)\nonumber\\
    &&\quad\qquad+(-1)^n\binom{n+\alpha}{n}(\sin^{2n}(x/2)-1)\nonumber\:.
\end{eqnarray}

Finally, by using the identity
\begin{equation}
    \sin^{2n}(x/2)-1 = -\cos^2(x/2) \sum_{j=1}^n \sin^{2(n-j)}(x/2),
\end{equation}

we prove the property \eqref{eq:10}
\begin{eqnarray}
    \widetilde{P}_n^{(\alpha,\beta)}(\cos x) = f_c(x) Q^{(\alpha,\beta)}_n(\cos(x))\:,
\end{eqnarray}
where
\begin{eqnarray}
    &&Q_n^{(\alpha,\beta)}(\cos (x))\\
    &&= \sum_{j=1}^n \bigg[(-1)^{n-j}\binom{n+\alpha}{j}\binom{n+\beta}{n-j}\cos^{2(j-1)}(x/2)\nonumber\\
    && \qquad\qquad+(-1)^n\binom{n + \beta}{n}\bigg]\sin^{2(n-j)}(x/2)\nonumber\:,
\end{eqnarray}

and $f_c(x) = \cos^2(x/2) = (1 + \cos(x))/2$. 
\bibliographystyle{unsrt}

\end{document}

\title{Cluster expansion constructed over Jacobi-Legendre polynomials for accurate force fields\\\phantom{}\\Supplementary Material}
\author{M.~Domina*}
\author{U.~Patil*}
\author{M.~Cobelli*}
\author{S.~Sanvito}
\affiliation{School of Physics and CRANN Institute, Trinity College Dublin, Ireland}
\maketitle

\section{Linear scaling}
\subsection{Three-body}
In this section we will prove that the two expressions
\begin{equation}\label{eq:1}
    \varepsilon_i^{(3)} = \sum_{(j,k)_i}\sum_{n_1 n_2 l}^{\text{unique}} a_{n_1 n_2 l}^{Z_j Z_k Z_i} \sum_{\text{symm.}}\bigg({\overline{P}}_{n_1 ji}^{(\alpha,\beta)}\overline{P}_{n_2ki}^{(\alpha,\beta)}P_l^{jki}\bigg),
\end{equation}
and
\begin{equation}\label{eq:2}
    \varepsilon_i^{(3)} = \sum_{\substack{Z_1Z_2\\Z_1\ge Z_2}} \sum_{n_1 n_2 l}^{\text{unique}} b_{n_1 n_2 l}^{Z_1 Z_2 Z_i}\bigg[C^{(3),Z_1 Z_2}_{i n_1 n_2 l} - S^{(3),Z_1 Z_2}_{i n_1 n_2}\bigg],
\end{equation}
where the coupling term $C^{(3),Z_1 Z_2}_{i n_1 n_2 l}$ and the self energy term $S^{(3),Z_1 Z_2}_{i n_1 n_2}$ are written in terms of the JL-atomic basis
\begin{equation}
        (J_p L_q)^{i,Z}_{n_1\ldots n_p l_1 m_1 \ldots l_q m_q} = \sum_{j\in Z}\bigg[\prod_{r=1}^p \overline{P}_{n_r ji}^{(\alpha,\beta)}\bigg] \bigg[\prod_{s=1}^q Y_{l_s}^{m_s}(\hat{\bm r}_{ji})\bigg]\:,
\end{equation}
as
\begin{equation}\label{eq:3}
\begin{dcases}
C^{(3),Z_1 Z_2}_{i n_1 n_2 l} = \dfrac{4\pi}{2l+1} \sum_{m=-l}^l (-1)^m (J_1L_1)^{i,Z_1}_{\substack{n_1 l m}} (J_1L_1)^{i,Z_2}_{\substack{n_2 l-m}},\\
S^{(3),Z_1 Z_2}_{in_1 n_2} = \delta_{Z_1 Z_2} (J_2L_0)^{i,Z_1}_{\substack{n_1 n_2}},
\end{dcases}
\end{equation}
are equivalent. The relation between the coefficients $a$ and $b$ can be found in eq. \eqref{eq:9}. 

Starting from \eqref{eq:1}, we have that the sum on $(j,k)_i$, namely all the proper pairs of atoms in the neighborhood of the i-th atom, can be re-arranged so that we always consider atoms such that their atomic number is in decreasing order. Therefore we can have the ordering

\begin{equation}\label{eq:5}
    \sum_{(j,k)_i} = \sum_{\substack{Z_1,Z_2\\Z_1\ge Z_2}}\sum_{\substack{(j,k)_i\\j\in Z_1, k\in Z_2}},
\end{equation}

and we can now separate in cases with atoms belonging to same species or to different ones. For the latter, we have that no symmetries among the coefficients is present, and therefore

\begin{equation}\label{eq:6}
     \sum_{n_1 n_2 l}^{\text{unique}}a_{n_1 n_2 l}^{Z_1 Z_2 Z_i} \sum_{\text{symm.}}\bigg({\overline{P}}_{n_1 ji}^{(\alpha,\beta)}\overline{P}_{n_2ki}^{(\alpha,\beta)}P_l^{jki}\bigg) = \sum_{n_1 n_2 l}^{\text{unique}}a_{n_1 n_2 l}^{Z_1 Z_2 Z_i} {\overline{P}}_{n_1 ji}^{(\alpha,\beta)}\overline{P}_{n_2ki}^{(\alpha,\beta)}P_l^{jki},
\end{equation}

where the sum here is not constrained by any symmetries. Here the atoms j and the atom k are necessarily different, belonging to different atomic species. Thus in this case we can freely sum all the atoms in the environment, so that, from the re-arrangement \eqref{eq:5}, we get
\begin{equation}\label{eq:7}
        \varepsilon_i^{(3)} =  \sum_{\substack{Z_1,Z_2\\Z_1 > Z_2}}\sum_{\substack{j\in Z_1 \\k\in Z_2}}\sum_{n_1 n_2 l}^{\text{unique}} a_{n_1 n_2 l}^{Z_1 Z_2 Z_i} {\overline{P}}_{n_1 ji}^{(\alpha,\beta)}\overline{P}_{n_2ki}^{(\alpha,\beta)}P_l^{jki}+\sum_{Z_1}\sum_{\substack{(j,k)_i\\j \in Z_1\\k\in Z_2}} \sum_{n_1 n_2 l}^{\text{unique}}a_{n_1 n_2 l}^{Z_1 Z_1 Z_i} \sum_{\text{symm.}}\bigg({\overline{P}}_{n_1 ji}^{(\alpha,\beta)}\overline{P}_{n_2ki}^{(\alpha,\beta)}P_l^{jki}\bigg).
\end{equation}

Dealing with the same species case, the second addend of the above equation, we have

\begin{eqnarray}\label{eq:8}
    &&\sum_{Z_1}\sum_{\substack{(j,k)_i\\j \in Z_1\\k\in Z_1}} \sum_{n_1 n_2 l}^{\text{unique}}a_{n_1 n_2 l}^{Z_1 Z_1 Z_i} \sum_{\text{symm.}}\bigg({\overline{P}}_{n_1 ji}^{(\alpha,\beta)}\overline{P}_{n_2ki}^{(\alpha,\beta)}P_l^{jki}\bigg)\nonumber \\
    &&= \sum_{Z_1}\sum_{\substack{(j,k)_i\\j \in Z_1\\k\in Z_1}} \Bigg[\sum_{\substack{n_1,n_2\\n_1 > n_2} l} a_{n_1 n_2 l}^{Z_1 Z_1 Z_i} \bigg({\overline{P}}_{n_1 ji}^{(\alpha,\beta)}\overline{P}_{n_2ki}^{(\alpha,\beta)}+{\overline{P}}_{n_2 ji}^{(\alpha,\beta)}\overline{P}_{n_1ki}^{(\alpha,\beta)}\bigg)P_l^{jki} + \sum_{n_1 l} a_{n_1 n_1 l}^{Z_1 Z_1 Z_i}{\overline{P}}_{n_1 ji}^{(\alpha,\beta)}\overline{P}_{n_1ki}^{(\alpha,\beta)}P_l^{jki}\Bigg].
\end{eqnarray}

The terms inside the curly bracket here can be seen as the swapping between j and k, so that we can un-restrict the sum on $(j,k)_i$ so that each proper pairs is counted once. However, the second addend, will be now be counted twice, therefore we can introduce the coefficients

\begin{equation}\label{eq:9}
    b_{n_1n_2l}^{Z_1 Z_2 Z_i} = \dfrac{a_{n_1n_2l}^{Z_1 Z_2 Z_i}}{1+\delta_{Z_1Z_2}\delta_{n_1 n_2}}\:,
\end{equation}
where we used the $\delta$-Kronecker to halve the coefficients when needed, so that the expression \eqref{eq:8} can be written as

\begin{equation}\label{eq:10}
    \eqref{eq:8} = \sum_{Z_1}\sum_{\substack{j \in Z_1\\k\in Z_1\\j \neq k}} \sum_{n_1,n_2 l}^\text{unique} b_{n_1 n_2 l}^{Z_1 Z_1 Z_i} {\overline{P}}_{n_1 ji}^{(\alpha,\beta)}\overline{P}_{n_2ki}^{(\alpha,\beta)}P_l^{jki}.
\end{equation}        
We now remove the restriction $j\neq k$ by adding and subtracting the term $j=k$, so that the expression becomes
\begin{equation}
    \eqref{eq:10} = \sum_{Z_1} \sum_{n_1 n_2 l}^\text{unique} b_{n_1 n_2 l}^{Z_1 Z_1 Z_i} \left[\sum_{\substack{j \in Z_1\\k\in Z_1}}{\overline{P}}_{n_1 ji}^{(\alpha,\beta)}\overline{P}_{n_2ki}^{(\alpha,\beta)}P_l^{jki} - \sum_{j\in Z_1} {\overline{P}}_{n_1 ji}^{(\alpha,\beta)}{\overline{P}}_{n_2 ji}^{(\alpha,\beta)}\right].
\end{equation}

Plugging this expression back in \eqref{eq:7}, and using the definition of the $b$ coefficients \eqref{eq:9}, we get

\begin{equation}\label{eq:12}
            \varepsilon_i^{(3)} =  \sum_{\substack{Z_1,Z_2\\Z_1 \geq Z_2}}\sum_{n_1 n_2 l}^{\text{unique}} b_{n_1 n_2 l}^{Z_1 Z_2 Z_i}\left[\sum_{\substack{j\in Z_1 \\k\in Z_2}} {\overline{P}}_{n_1 ji}^{(\alpha,\beta)}\overline{P}_{n_2ki}^{(\alpha,\beta)}P_l^{jki}-\delta_{Z_1Z_2}\sum_{j\in Z_1}{\overline{P}}_{n_1 ji}^{(\alpha,\beta)}\overline{P}_{n_2ji}^{(\alpha,\beta)}\right].
\end{equation}

Finally, by using the addition theorem of the spherical harmonics \cite{Angular}
\begin{equation}\label{eq:13}
    P_l^{jki} = P_l(\hat{\bm{r}}_{ji}\cdot \hat{\bm{r}}_{ki}) = \dfrac{4\pi}{2l+1} \sum_{m=-l}^l (-1)^m Y^m_l(\hat{\bm{r}}_{ji}) Y^{-m}_l(\hat{\bm{r}}_{ki}),
\end{equation}
we can separate the single contributions of \eqref{eq:12} as

\begin{equation}
     \varepsilon_i^{(3)} =  \sum_{\substack{Z_1,Z_2\\Z_1 \geq Z_2}}\sum_{n_1 n_2 l}^{\text{unique}} b_{n_1 n_2 l}^{Z_1 Z_2 Z_i}\left[\dfrac{4\pi}{2l+1}\sum_{m=-l}^m(-1)^m\left(\sum_{j\in Z_1}{\overline{P}}_{n_1 ji}^{(\alpha,\beta)}Y^m_l(\hat{\bm{r}}_{ji})\right)\left(\sum_{k\in Z_2} \overline{P}_{n_2ki}^{(\alpha,\beta)}Y^{-m}_l(\hat{\bm{r}}_{ki})\right)-\delta_{Z_1Z_2}\sum_{j\in Z_1}{\overline{P}}_{n_1 ji}^{(\alpha,\beta)}\overline{P}_{n_2ji}^{(\alpha,\beta)}\right],
\end{equation}

so that, using the definition of the JL-atomic basis (\eqref{eq:3})
\begin{equation}
    \begin{dcases}
    (J_1L_1)^{i,Z}_{nlm} = \sum_{j\in Z}{\overline{P}}_{n_1 ji}^{(\alpha,\beta)}Y^m_l(\hat{\bm{r}}_{ji}),\\
    (J_2L_0)^{i,Z}_{n_1 n_2} = \sum_{j\in Z}{\overline{P}}_{n_1 ji}^{(\alpha,\beta)}\overline{P}_{n_2ji}^{(\alpha,\beta)}\:,
    \end{dcases}
\end{equation}
ends the proof.

We remark that $(J_1L_1)^{i,Z}_{nlm}$ is equivalent to the atomic basis presented in ACE \cite{ACE}, $A_{inlm}^Z$. Also the $(J_2L_0)^{i,Z}_{n_1 n_2}$ can be written in terms of the same atomic basis. Indeed, we have
\begin{equation}
    {\overline{P}}_{n_1 ji}^{(\alpha,\beta)}\overline{P}_{n_2ji}^{(\alpha,\beta)} = \sum_{n=2}c_n^{n_1n_2}{\overline{P}}_{n ji}^{(\alpha,\beta)}\:,
\end{equation}
since the Jacobi polynomials are complete. We use the double-vanishing Jacobi polynomials in the expansion since product on the LHS still vanishes at both the origin and the cut-off radius. The coefficients $c_n^{n_1n_2}$ can be evaluated by mean of the integral

\begin{equation}
    c_n^{n_1n_2} = A^{(\alpha,\beta)}_{n}\int_{-1}^{1}  (1-x)^\alpha (1+x)^\beta  {\overline{P}}_{n_1}^{(\alpha,\beta)}(x)\overline{P}_{n_2}^{(\alpha,\beta)}(x) P_{n}^{(\alpha,\beta)}(x)\dd x\:,
\end{equation}
where the normalization constant $A^{(\alpha,\beta)}_{n}$ is \cite{Abramovitz}
\begin{equation*}
A_n^{(\alpha,\beta)} = \dfrac{2n+\alpha+\beta+1}{2^{\alpha+\beta+1}}\dfrac{\Gamma(n+\alpha+\beta+1)n!}{\Gamma(n+\alpha+1)\Gamma(n+\beta+1)},
\end{equation*}
being $\Gamma(x)$ the Gamma function. We then deduce that we can write
\begin{equation}\label{eq:18}
    (J_2L_0)^{i,Z}_{n_1 n_2} = \sum_{n=2}c_n^{n_1n_2}(J_1L_1)^{i,Z}_{nl0} = \sum_{n=2} c_n^{n_1n_2}A^{Z}_{inl0}\:.
\end{equation}
However evaluating the coefficients $c_n^{n_1n_2}$ is not more efficient than simply perform the products of polynomials and keep more indexes. Also, by mean of eq. \eqref{eq:18}, we can say the self energy is effectively a 2B term and, as such, could be re-absorbed in lower order terms. However, not only the 2B cases do not necessarily share the same basis (the implied hyperparameters $\alpha$, $\beta$ and the cut-off could all be different), but also, the maximum degree should be much larger than the one of the 3B cases. For these reasons, we introduced the JL-atomic basis in the first place, and we argue that the self-energy should be considered separately and subtracted explicitly. In this way, also the equivalence proved in this section is preserved, so that one could jump from the JL-atomic basis representation to the internal coordinate expansion at any moment, e.g. if the number of atoms in the neighborhood is small and, as such, the linear scaling is unnecessary.

\subsection{Four-body}
For the 4B local energy, $\varepsilon_i^{(4)}$, the equivalent expressions that one can use to go from the internal coordinate representation to the JL-atomic basis are
\begin{equation}\label{eq:19}
    \varepsilon_i^{(4)}= \sum_{(j,k,p)_i}\sum_{\substack{n_1 n_2 n_3\\l_1 l_2 l_3}}^{\text{unique}} a_{\substack{n_1 n_2 n_3\\l_1 l_2 l_3 }}^{Z_jZ_kZ_pZ_i} \sum_{\text{symm.}}\bigg({\overline{P}}_{n_1 ji}^{(\alpha,\beta)}\overline{P}_{n_2ki}^{(\alpha,\beta)}\overline{P}_{n_3pi}^{(\alpha,\beta)}P_{l_1}^{jki}P_{l_2}^{jpi}P_{l_3}^{kpi}\bigg)\:,
\end{equation}
and 
\begin{equation}\label{eq:20}
     \varepsilon_i^{(4)}=\sum_{Z_1\ge Z_2\ge Z_3} \sum_{\substack{n_1 n_2 n_3\\ l_1 l_2 l_3}}^{\text{unique}} b_{\substack{n_1 n_2 n_3\\ l_1l_2l_3}}^{Z_1 Z_2 Z_3 Z_i}\Bigg[C^{(4),Z_1 Z_2 Z_3}_{i ,\substack{n_1 n_2 n_3\\ l_1l_2l_3}} - S^{(4),Z_1 Z_2 Z_3}_{i ,\substack{n_1 n_2 n_3\\ l_1l_2l_3}}\Bigg]\:,
\end{equation}
where relation between coefficients is
\begin{equation}\label{eq:21}
    b_{\substack{n_1 n_2 n_3\\ l_1l_2l_3}}^{Z_1 Z_2 Z_3 Z_i} = \bigg[1+\delta_{Z_1Z_2}\delta_{n_1 n_2}\delta_{l_2l_3}+\delta_{Z_2Z_3}\delta_{n_2 n_3}\delta_{l_1l_2}+\delta_{Z_1Z_2}\delta_{Z_2Z_3}\big(\delta_{n_1n_3}\delta_{l_1l_3}+2\delta_{n_1n_3}\delta_{n_2n_3}\delta_{l_1l_3}\delta_{l_1l_2}\big)\bigg]^{-1}a_{\substack{n_1 n_2 n_3\\l_1 l_2 l_3 }}^{Z_1Z_2Z_3Z_i}\:,
\end{equation}
which takes care of the division by 2 or by 6 when necessary (to properly take into account the symmetric cases), the coupling term is
\begin{equation}\label{eq:22}
C^{(4),Z_1 Z_2 Z_3}_{i ,\substack{n_1 n_2 n_3\\ l_1l_2l_3}} = \dfrac{(4\pi)^3}{(2l_1+1)(2l_2+1)(2l_3+1)}\sum_{m_1 m_2 m_3}(-1)^{m_1 + m_2 + m_3}(J_1L_2)^{i,Z_1}_{\substack{n_1 l_1 m_1 l_2 -m_2 }}(J_1L_2)^{i,Z_2}_{\substack{n_2 l_3 m_3 l_1 -m_1 }}(J_1L_2)^{i,Z_3}_{\substack{n_3 l_2 m_2 l_3 -m_3 }}\:,
\end{equation}
and the self energy term is
\begin{eqnarray}\label{eq:23}
    S^{(4),Z_1 Z_2 Z_3}_{i ,\substack{n_1 n_2 n_3\\ l_1l_2l_3}} &=&\delta_{Z_1Z_2}\dfrac{(4\pi)^2}{(2l_2+1)(2l_3+1)}\sum_{m_2m_3}(-1)^{m_2+m_3} (J_2L_2)^{i,Z_1}_{n_1 n_2 l_2 m_2 l_3 m_3 }(J_1L_2)^{i,Z_3}_{n_3 l_2 -m_2 l_3 -m_3 }\nonumber\\
    &&+\delta_{Z_2Z_3}\dfrac{(4\pi)^2}{(2l_1+1)(2l_2+1)}\sum_{m_1m_2}(-1)^{m_1+m_2} (J_2L_2)^{i,Z_2}_{n_2 n_3 l_1 m_1 l_2 m_2 }(J_1L_2)^{i,Z_1}_{n_1 l_1 -m_1 l_2 -m_2 }\\
    &&+\delta_{Z_1Z_2}\delta_{Z_2Z_3}\left[\dfrac{(4\pi)^2}{(2l_1+1)(2l_3+1)}\sum_{m_1m_3}(-1)^{m_1+m_3} (J_2L_2)^{i,Z_1}_{n_1 n_3 l_1 m_1 l_3 m_3 }(J_1L_2)^{i,Z_1}_{n_2 l_1 -m_1 l_3 -m_3 }+(J_3L_0)^{i,Z_1}_{n_1n_2n_3}\right]\nonumber\:.
\end{eqnarray}

The derivation of the equivalence between \eqref{eq:19} and \eqref{eq:20}, closely follows the one for the 3B case. The only remark is that here we choose the arrangement 
\begin{equation}
    \sum_{(j,k,p)_i} = \sum_{\substack{Z_1,Z_2,Z_3\\Z_1\ge Z_2\ge Z_3}}\sum_{\substack{(j,k,p)_i\\j\in Z_1, k\in Z_2,p \in Z_3}},
\end{equation}
implying that one have to consider explicitly when in presence of atoms of the same species. This gives raise to all the possible cases that are present in the coefficients relation \eqref{eq:21} and in the self energy \eqref{eq:23}. The coupling term will be compared to the one proposed for the Bispectrum components \cite{Bartok}, and so it is useful to explicitly show that, once the rearrangement is made and the constraints on the summations over the atoms in the neighborhood are relieved, one obtains
 
 \begin{equation}\label{eq:25}
     C^{(4),Z_1 Z_2 Z_3}_{i ,\substack{n_1 n_2 n_3\\ l_1l_2l_3}} = \sum_{\substack{j\in Z_1\\k\in Z_2\\ p\in Z_3}}{\overline{P}}_{n_1 ji}^{(\alpha,\beta)}\overline{P}_{n_2ki}^{(\alpha,\beta)}\overline{P}_{n_3pi}^{(\alpha,\beta)}P_{l_1}^{jki}P_{l_2}^{jpi}P_{l_3}^{kpi}\:.
 \end{equation}
Now, by applying the addition theorem for spherical harmonics \eqref{eq:13} three times, and by separating the contributions for each atoms, we get \eqref{eq:22}. From here, it is clear why we use the terms $(J_1L_2)^{i,Z}$: indeed, for each atoms, the expression presents one double-vanishing Jacobi polynomials and two Legendre polynomials. Moreover, by using the property

\begin{equation}\label{eq:26}
Y^{m_1}_{l_1}(\hat{\bm r})Y^{m_2}_{l_2}(\hat{\bm r})=\sum_{lm} (-1)^m \sqrt{\dfrac{(2l+1)(2l_1+1)(2l_2+1)}{4\pi}}\\
\mqty(l_1 &l_2 &l\\0 &0& 0)\mqty(l_1 &l_2 &l\\m_1 &m_2& -m)  Y^{m}_{l}(\hat{\bm r})\:,
\end{equation}
for the product of two spherical harmonics (with the use of the 3j-Wigner symbols \cite{Angular}), one can deduce the analogous relation between $(J_1L_2)_{nl_1m_1l_2m_2}^{i,Z}$ and the atomic basis $A_{inlm}^Z$
\begin{equation}
    (J_1L_2)_{nl_1m_1l_2m_2}^{i,Z}=\sum_{lm} (-1)^m \sqrt{\dfrac{(2l+1)(2l_1+1)(2l_2+1)}{4\pi}}\\
\mqty(l_1 &l_2 &l\\0 &0& 0)\mqty(l_1 &l_2 &l\\m_1 &m_2& -m)A_{inlm}^Z\:.
\end{equation}

\section{Connecting the internal coordinate representation with the Bispectrum components}

In this section, we report how the 4B coupling term, $C^{(4),Z_1 Z_2 Z_3}_{i ,\substack{n_1 n_2 n_3\\ l_1l_2l_3}}$, can be written as a linear combination of bispectrum components. The expression shows how to sum bispectrum components so that the result is casted in terms of internal coordinate only (resulting in the coupling term). We note that, while the bispectrum components are more general, because they can distinguish reflections, the internal-coordinate representation lacks this property. 

 Writing the bispectrum components in the ACE formalism, that is, by using the 3j-Wigner symbols, as

\begin{equation}
    B_{i\substack{n_1n_2n_3\\l_1 l_2 l_3}}^{(3),Z_1 Z_2 Z_3} =\sum_{m_1 m_2 m_3}\mqty(l_1 & l_2 & l_3\\m_1 &m_2& m_3) A^{Z_1}_{in_1 l_1 m_1}A^{Z_2}_{in_2 l_2 m_2}A^{Z_3}_{in_3 l_3 m_3},
\end{equation}

it holds that 

\begin{eqnarray}
    &&C^{(4),Z_1 Z_2 Z_3}_{i ,\substack{n_1 n_2 n_3\\ l_1l_2l_3}} = \dfrac{(4\pi)^3}{(2l_1+1)(2l_2+1)(2l_3+1)} \sum_{L_1 L_2 L_3} w_{l_1 l_2 l_3}^{L_1 L_2 L_3}  B_{i\substack{n_1n_2n_3\\L_1 L_2 L_3}}^{(3),Z_1 Z_2 Z_3}, 
\end{eqnarray}
with the expansion coefficients, $w_{l_1 l_2 l_3}^{L_1 L_2 L_3}$, defined as
\begin{equation}
    w_{l_1 l_2 l_3}^{L_1 L_2 L_3} = (-1)^{L_1 + L_2 + L_3 + l_1 + l_2 + l_3}\mqty(l_1 & l_2 & L_1\\0 & 0 & 0)\mqty(l_3 & l_1 & L_2\\0 & 0 & 0)\mqty(l_2 & l_3 & L_3\\0 & 0 & 0)\sj{L_3}{L_2}{L_1}{l_1}{l_2}{l_3}.
\end{equation}

The expansion coefficients are here written in terms of 3j- and 6j-Wigner symbols \cite{Angular}. Given eq.~\eqref{eq:25}, this expression can be proven by expanding the Legendre polynomials as in eq.~\eqref{eq:13}, contracting spherical harmonics with the same argument as in eq.\eqref{eq:26}, and using the contraction rule for the 3j-Wigner symbols given in \url{http://functions.wolfram.com/07.39.23.0012.01}. Finally rearranging all the term using the symmetry properties of the 3j-Wigner symbols and the complex conjugate of spherical harmonics, leads directly to the expressions above.

\section{Fit on the Carbon dataset}
\begin{figure*}
\centering
\includegraphics[width=\textwidth]{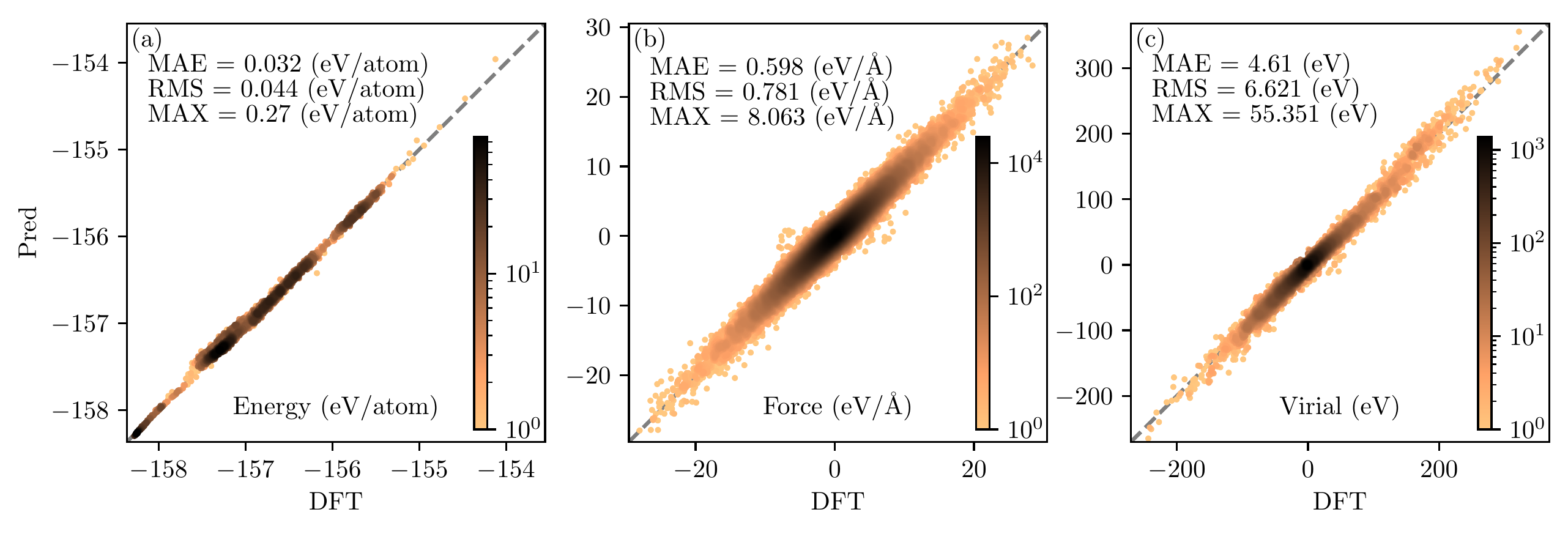}
\caption{Parity plots on the training set for (a) energies, (b) forces, (c) virial stress. }
\label{fig:train}
\end{figure*}

\begin{figure*}
\centering
\includegraphics[width=\textwidth]{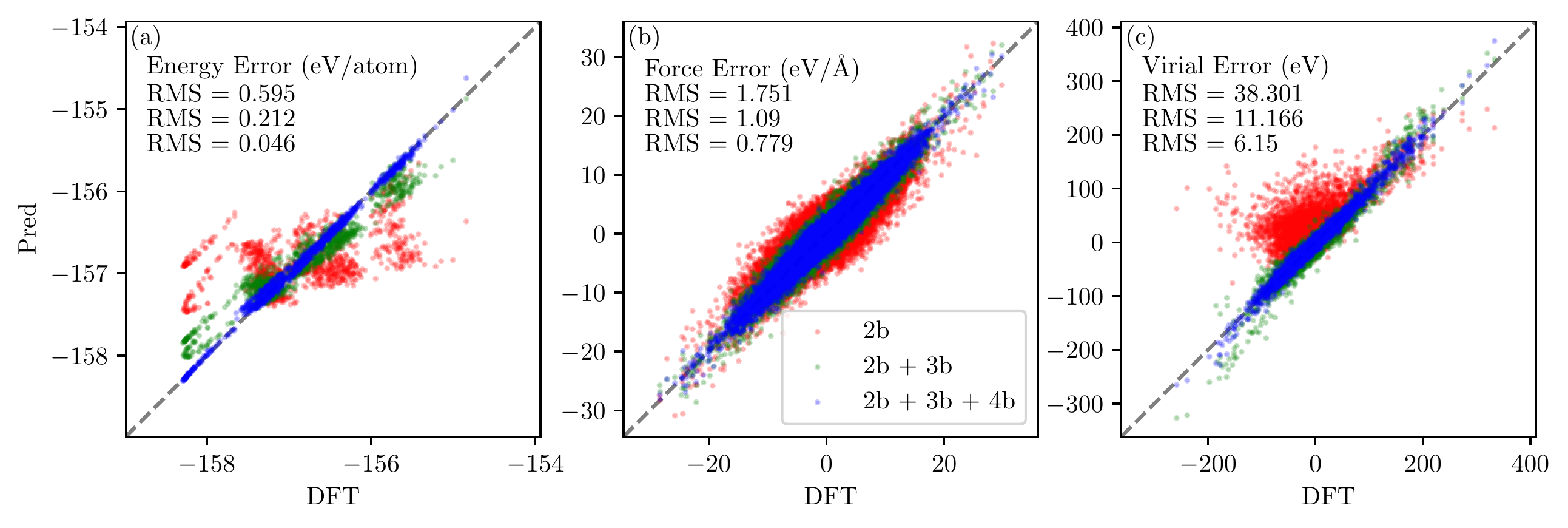}
\caption{Parity plots on the test set for (a) energies, (b) forces, (c) virial stress with increasing body order. The RMSE are reported for each plot. }
\label{fig:hierarchy}
\end{figure*}

We report here, in Fig.~\ref{fig:train}, the parity plots on the prediction set for the energies, the forces and the stress. In Fig.~\ref{fig:hierarchy}, we also reported a hierarchical study showing the improvement with increasing body order of the fit. There, it can be appreciated how the 2B fit performs very badly on the energy, as expected from the reduced number of features, while the 3B significantly improve the fit. However it still does not discriminate well between two crystalline phases present in the dataset, for which we need the 4B. Also, while the fit on the forces performs unexpectedly well already at the 2B level, given just the 11 features used, it still requires higher body order to reach satisfactory accuracy levels. The hyperparameters for this two fits (2B and 2B+3B) are the same of the one reported in the main text.

\FloatBarrier